\documentclass[graybox]{svmult}

\usepackage{mathptmx}       % selects Times Roman as basic font
\usepackage{helvet}         % selects Helvetica as sans-serif font
\usepackage{courier}        % selects Courier as typewriter font
\usepackage{makeidx}         % allows index generation
\usepackage{graphicx}        % standard LaTeX graphics tool
\usepackage{multicol}        % used for the two-column index
\usepackage[bottom]{footmisc}% places footnotes at page bottom

\usepackage{amssymb}
\usepackage{amsmath}

\usepackage[ruled,vlined]{algorithm2e}
\usepackage{algorithmic}

\newcommand{\K}{\mathbb{K}}

\newcommand{\bi}{\begin{itemize}}
\newcommand{\ei}{\end{itemize}}
\newcommand{\be}{\begin{enumerate}}
\newcommand{\ee}{\end{enumerate}}

\usepackage{fca}

\makeindex             % used for the subject index
\begin{document}

\title*{Multimodal Clustering for Community Detection}
\author{Dmitry~I.~Ignatov\inst{1}  \and Alexander~Semenov\inst{1,2} \and Daria~Komissarova\inst{1} \and Dmitry~V.~Gnatyshak\inst{1}}
\institute{Dmitry~I.~Ignatov \at \inst{1}National Research University Higher School of Economics, Moscow, Russia, \email{dignatov@hse.ru}
\and Alexander~Semenov \at \inst{1}National Research University Higher School of Economics, Moscow, Russia, and ~\inst{2}Mobile TeleSystems PJSC, Moscow, Russia, \email{SemenoffAlex@gmail.com }
\and Daria~Komissarova \at \inst{1}National Research University Higher School of Economics, Moscow, Russia,  \email{komissarovadaria93@gmail.com}
\and Dmitry~V.~Gnatyshak \at \inst{1}National Research University Higher School of Economics, Moscow, Russia,, \email{dgnatyshak@hse.ru}
}
\maketitle

\abstract*{Multimodal clustering is an unsupervised technique for mining interesting patterns in $n$-adic binary relations or $n$-mode networks. Among different types of such generalized patterns one can find biclusters and formal concepts (maximal bicliques) for 2-mode case, triclusters and triconcepts for 3-mode case, closed $n$-sets for $n$-mode case, etc. Object-attribute biclustering (OA-biclustering) for mining large binary datatables (formal contexts or 2-mode networks) arose by the end of the last decade due to intractability of computation problems related to  formal concepts; this type of patterns was proposed as a meaningful and scalable approximation of formal concepts. 
In this paper, our aim is to present recent advance in OA-biclustering and its extensions to mining multi-mode communities in SNA setting. We also discuss connection between clustering coefficients known in SNA community for 1-mode and 2-mode networks and OA-bicluster density, the main quality measure of an OA-bicluster. Our experiments with 2-, 3-, and 4-mode large real-world networks show that this type of patterns is suitable for community detection in multi-mode cases within reasonable time even though the number of corresponding $n$-cliques is still unknown due to computation difficulties. An interpretation of OA-biclusters for 1-mode networks is provided as well.  
.}

\abstract{Multimodal clustering is an unsupervised technique for mining interesting patterns in $n$-adic binary relations or $n$-mode networks. Among different types of such generalized patterns one can find biclusters and formal concepts (maximal bicliques) for 2-mode case, triclusters and triconcepts for 3-mode case, closed $n$-sets for $n$-mode case, etc. Object-attribute biclustering (OA-biclustering) for mining large binary datatables (formal contexts or 2-mode networks) arose by the end of the last decade due to intractability of computation problems related to  formal concepts; this type of patterns was proposed as a meaningful and scalable approximation of formal concepts. 
In this paper, our aim is to present recent advance in OA-biclustering and its extensions to mining multi-mode communities in SNA setting. We also discuss connection between clustering coefficients known in SNA community for 1-mode and 2-mode networks and OA-bicluster density, the main quality measure of an OA-bicluster. Our experiments with 2-, 3-, and 4-mode large real-world networks show that this type of patterns is suitable for community detection in multi-mode cases within reasonable time even though the number of corresponding $n$-cliques is still unknown due to computation difficulties. An interpretation of OA-biclusters for 1-mode networks is provided as well.  
}


\section{Introduction}

 Online social networking services generate massive amounts of data, which can become a valuable source for guiding Internet advertisement efforts or provide sociological insights. Each registered user has a network of friends as well as specific profile features. These profile features describe the user's tastes, preferences, the groups he or she belongs to, etc. Social Network Analysis (SNA) is a popular research field in which methods are developed for analysing 1-mode networks, like friend-to-friend\footnote{\url{https://en.wikipedia.org/wiki/Friend-to-friend}}, 2-mode or affilliation networks \cite{Latapy:2008,Liu:2010,Opsahl:2011}, 3-mode \cite{Fararo:1984,Jaschke:2006,Murata:2010,Ignatov:2011,Bohman:2012} and even multimode dynamic networks \cite{Roth:2005,Tang:2008,Roth:2010,Yavorsky:2011}. By multimode networks we mean namely such networks where actors can be related with other types of entities by edges like those between users and their interests in two-mode case or by hyperedges like those related users, tags, and resources in three-mode case; sometimes such networks are called heterogeneous since different types of nodes are involved \cite{Jones:2015}.  We focus on the subfield of bicommunity identification and its higher order extenstions. Thus, in particular, we present tri- and tetracommunities examples extracted from real data. For one-mode case a reader may refer to an extensive survey on community detection \cite{Fortunato:2010}.

The notion of community in SNA and Complex Networks is closely related to the notion of cluster in Data Analysis \cite{Fortunato:2010,Barabasi:2016}. There is the main issue in both disciplines:  what is a common definition of community and what is a common definition of cluster? On the one hand, it is clear that actors from the same community should be similar as well as objects in one cluster; on the other hand, these actors (or objects) should be less similar to actors (or objects) from another community (or cluster). This general idea allows a variety of definitions suitable for concrete purposes in both domains \cite{Fortunato:2010,Barabasi:2016,Mirkin:1996}.

There is a large amount of network data that can be represented as bipartite or tripartite graphs.
Standard techniques for community detection in two-mode networks like ``maximal bicliques search'' return a huge number of patterns (in the worst case exponential w.r.t. the input size)~\cite{Roth:2006,Kuznetsov:2007a}. Moreover, not all members of such bicommunites should be related to the same items, for example,  exactly the same vocabulary used by each member in case of epistemic communities. 
Therefore we need some relaxation of the biclique notion as well as appropriate interestingness measures and constraints for mining and filtering such ``relaxed'' biclique communities.

Applied lattice theory provides us with a notion of formal concept \cite{Ganter:1999:FCA}, which is identical to biclique; formal concepts and  concept lattices (or Galois lattices) are widely known in the social network analysis community (see, e.g. \cite{Freeman:1993,Freeman:1996,Duquenne:1996,White:1996,Mohr:1997,Roth:2006}). However, these methods are overly rigid for analysing large amounts of data resulting in a huge number of concepts even if their computation is feasible. 

A concept-based bicluster (or object-attribute bicluster) \cite{Ignatov:2010} is a scalable approximation of a formal concept (biclique). The advantages of concept-based biclustering are:

\begin{enumerate}

 \item Less number of patterns to analyse (no more than the number of edges in the original network);
 \item Less computational time (polynomial vs exponential);
 \item Tolerance to missing (object, attribute) pairs;
 \item Filtering of biclusters (communites) by density threshold.
\end{enumerate}

In general, the method of biclustering dates back to the seminal work of Hartgian on the so-called direct clustering~\cite{Hartigan:1972}, where clusters of objects may appear sharing only a subset of attributes. The term biclustering was introduced later in the book of Mirkin~\cite{Mirkin:1996}:
\begin{quote}
The term biclustering refers to simultaneous clustering of both row and column sets
in a data matrix. Biclustering addresses the problems of aggregate representation
of the basic features of interrelation between rows and columns as expressed in the
data.
\end{quote}

Following this terminology, formal concepts can be considered as maximal inclusion biclusters of constant values in binary data~\cite{Kaytoue:2011}, whereas their relaxations tolerant to missing object-attribute pairs can be called object-attribute biclusters~\cite{Ignatov:2010,Ignatov:2012a}. 

There are several sucessful attempts to mine 2-mode \cite{Roth:2008,Krasnov:2014}, 3-mode \cite{Jaschke:2006}, and even 4-mode communities \cite{Jelassi:2015} by means of Formal Concept Analysis. For analysing three-mode network data like folksonomies \cite{Wal:2007} we have also proposed a scalable triclustering technique \cite{Ignatov:2013,Ignatov:2015}.

These studies for higher-mode cases were enabled by the previous introduction of the so-called triconcepts by Lehman and Wille \cite{Lehmann:1995,Wille:1995}; a formal triconcept consists of three components: extent (objects), intent (attributes),  and modus (conditions under which an object has an attribute). It is a matter of curiosity, but such triconcepts had been used  for analysing triadic data in social cognition studies~\cite{Ganter:1994} before their formal introduction. Later, a polyadic (or multimodal) extension of FCA was introduced in~\cite{Voutsadakis:02}.

Previously, we have introduced a pseudo-triclustering technique for tagging groups of users by their common interests~\cite{Gnatyshak:2012}. This approach differs from traditional triclustering methods because it relies on the extraction of biclusters from two separate object-attribute tables and belongs rather to methods for analysing multi-relational networks.  Here we investigate applicability of biclustering and triclustering (as well as $n$-clustering, its higher-mode extension) to community detection in two-, three- and higher-mode networks directly.

The remainder of the paper is organized as follows. In Section~\ref{def}, we introduce basic notions of Formal Concept Analysis. Section~\ref{nclust} describes object-attribute biclustering and its direct generalisations to higher dimensions. Section~\ref{sec:quality} briefly discuss a variety of quality measures used in clustering, FCA, and SNA domains and their interrelation with multimodal clustering.  In Section~\ref{data}, we describe datasets which we  have chosen to illustrate the performance of the approach. We present the results obtained during experiments on these datasets in Section~\ref{exp}. Related work is discussed in Section~\ref{sec:rel}, while Section~\ref{con} concludes our paper and describes some interesting directions for future research.

\section{Basic definitions}\label{def}

\subsection{Formal Concept Analysis}\label{FCA}

\emph{A formal context} in FCA \cite{Ganter:1999:FCA} is a triple $\mathbb{K} = (G, M, I)$, where $G$ is a \emph{set of objects}, $M$ is a \emph{set of attributes}, and the relation $I \subseteq G \times M$ shows which object possesses which attribute. For any $A \subseteq G$ and $B \subseteq M$ one can define \emph{Galois operators}:

\begin{eqnarray}
A' = \{m\in M\mid gIm \ {\rm for\ all}\ g\in A\}, \\
B' = \{g\in G\mid gIm \ {\rm for\ all}\ m\in B\}. \nonumber
\end{eqnarray}

The operator $''$ (applying the operator $'$ twice) is a \emph{closure operator}: it is idempotent ($A'''' = A''$), monotone ($A \subseteq B$ implies $A'' \subseteq B''$) and extensive ($A \subseteq A''$). The set of objects $A \subseteq G$ such that $A'' = A$ is called closed. Similar properties are valid for closed attribute sets, subsets of a set $M$. A pair $(A, B)$ such that $A \subseteq G$, $B \subseteq M$, $A' = B$ and $B' = A$, is called a \emph{formal concept} of a context $\mathbb{K}$. The sets $A$ and $B$ are closed and called \emph{extent} and \emph{intent} of a formal concept $(A, B)$ correspondingly. For the set of objects $A$ the set of their common attributes $A'$ describes the similarity of objects of the set $A$, and the closed set $A''$ is a cluster of similar objects (with the set of common attributes $A'$). The relation ``to be a more general concept'' is defined as follows: $(A, B) \geq (C, D)$ iff $A\supseteq C$. The concepts of a formal context $\mathbb{K} = (G, M, I)$ ordered by extensions inclusion form a lattice, which is  called \emph{concept lattice}. For its visualization \emph{line diagrams} (Hasse diagrams) can be used, i.e. the cover graph of the relation ``to be a more general concept''.
In the worst case (Boolean lattice) the number of concepts is equal to $2^{\{\min{|G|,|M|\}}}$, thus, for large contexts, to make application of FCA machinery tractable the data should be sparse. Moreover, one can use different ways of filtering of formal concepts (for example, choosing concepts by their stability index or extent size).

\begin{figure}
\centering
\includegraphics[scale=0.65]{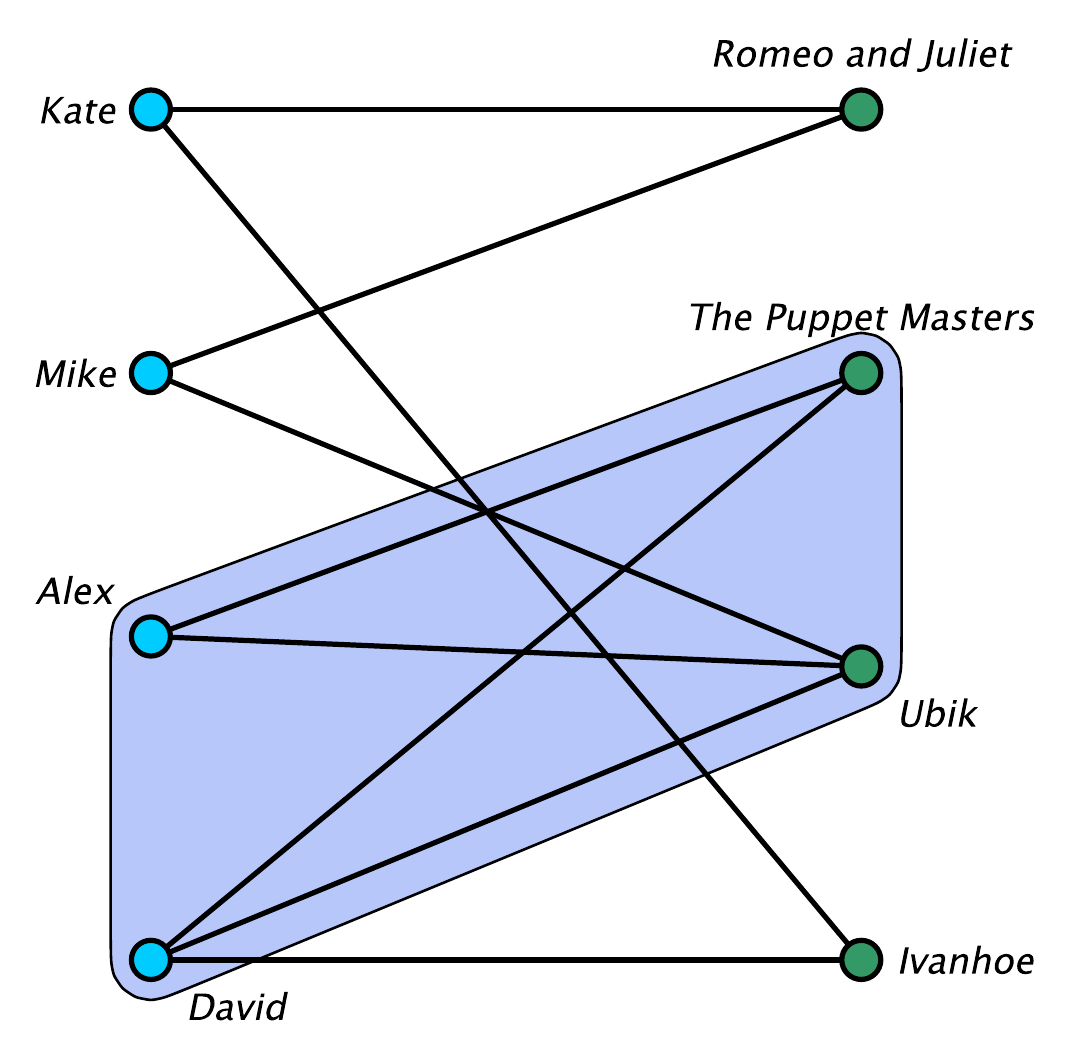}
\caption{Two-mode network of readers and its community of Sci-Fi readers (shaded)}\label{fig:readers}
\end{figure}

\begin{svgraybox}

Let us consider a formal context $\K$ that consists of four objects, persons (Alex, Mike, Kate, David), four attributes, books (Romeo and Juliet by William Shakespeare, The Puppet Masters by Robert A. Heinlein, Ubik by Philip K. Dick, and Ivanhoe by Walter Scott), and incidence relation showing which person which book read or liked.

\

\begin{center}
\begin{cxt}%
\cxtName{$\K$}%
\atr{$Romeo \ and \ Juliet$}%
\atr{$The \ Puppets \ Masters$}%
\atr{$Ubik$}%
\atr{$Ivanhoe$}%
\obj{x..x}{$Kate$} \obj{x.x.}{$Mike$} \obj{.xx.}{$Alex$} \obj{.xxx}{$David$}
\end{cxt}
\end{center}

\

There are nine concepts there. For example, 

$C_1=(\{Kate, Mike\}, \{Romeo \ and \ Juliet\})$

$C_2=(\{Alex, David\}, \{The \ Puppet \ Masters, Ubik\})$

$C_3=(\{Kate, David\}, \{Ivanhoe\})$.

Note that the pair of sets $(A,B)=(\{Alex, David\}, \{Ubik\})$ does not form a formal concept since we can enlarge its extent by one more object Mike to fulfil  $(A \cup \{Mike\})'=B$ and $B'=A\cup \{Mike\}$. So, $C_4=(\{MIke, Alex, David\}, \{Ubik\})$ is a formal concept. The corresponding bipartite graph is shown in Fig.~\ref{fig:readers} along with the biclique formed by elements of concept $C_2$.

\end{svgraybox}

From SNA viewpoint, if we assume that an OA-bicluster $(event',actor')$ is a community found, we are looking for a pair $(actor, event)$ in an input network, where this actor participated in all of the events typical for the community, while the chosen event is typical for all the members of that community.

\section{Higher-order extenstions of FCA and multimodal clustering}\label{nclust}

\subsection{Triadic and Polyadic FCA}

For convenience, a \emph{triadic context} is denoted by $(X_1,X_2,X_3,Y)$. A triadic context $\mathbb{K}=(X_1,X_2,X_3,Y)$ gives rise to the following dyadic contexts

$\mathbb{K}^{(1)}=(X_1, X_2\times X_3, Y^{(1)})$,
$\mathbb{K}^{(2)}=(X_2, X_1\times X_3, Y^{(2)})$,
$\mathbb{K}^{(3)}=(X_3, X_1\times X_2, Y^{(3)})$,

where $gY^{(1)}(m,b):\Leftrightarrow mY^{(2)}(g,b):\Leftrightarrow bY^{(3)}(g,m):\Leftrightarrow (g,m,b) \in Y$. The \emph{derivation operators} (primes or concept-forming operators) induced by $\mathbb{K}^{(i)}$ are denoted by $(.)^{(i)}$. For each induced dyadic context we have two kinds of such derivation operators. That is, for $\{i,j,k\}=\{1,2,3\}$ with $j<k$ and for $Z \subseteq X_i$ and $W \subseteq X_j\times X_k$, the $(i)$-derivation operators are defined by:

$$Z \mapsto Z^{(i)} = \{(x_j,x_k) \in X_j\times X_k| x_i, x_j, x_k \mbox{ are related by Y for all } x_i \in Z\},$$
$$W \mapsto W^{(i)} = \{x_i \in X_i| x_i, x_j, x_k \mbox{ are related by Y for all } (x_j,x_k) \in W\}.$$
Formally, a \emph{triadic concept} of a triadic context $\mathbb{K}=(X_1,X_2,X_3,Y)$ is a triple $(A_1,A_2,A_3)$ of $A_1 \subseteq X_1, A_2 \subseteq X_2, A_3 \subseteq X_3$, such that for every $\{i,j,k\}=\{1,2,3\}$ with $j<k$ we have $(A_j \times A_k)^{(i)}=A_i$.
For a certain triadic concept $(A_1,A_2,A_3)$, the components $A_1$, $A_2$, and $A_3$ are called the \emph{extent}, the \emph{intent}, and the \emph{modus} of $(A_1,A_2,A_3)$. Since a tricontext $\mathbb{K}=(X_1,X_2,X_3,Y)$ can be interpreted as a three-dimensional cross table, according to our definition, under suitable permutations of rows, columns, and layers of this cross table, the triadic concept $(A_1,A_2,A_3)$ is interpreted as a maximal cuboid full of crosses. The set of all triadic concepts of $\mathbb{K}=(X_1,X_2,X_3,Y)$ is denoted by $\mathfrak{T}(X_1,X_2,X_3,Y)$.

To avoid additional technical description of $n$-ary concept forming operators, we introduce $n$-adic formal concepts without their usage. The $n$-adic concepts of an $n$-adic context $(X_1,\ldots,X_n, Y)$ are exactly the maximal $n$-tuples $(A_1, \ldots , A_n)$ in $2^{X_1} \times \cdots \times 2^{X_n}$ with $A_1 \times \cdots \times A_n \subseteq Y$ with respect to component-wise set inclusion \cite{Voutsadakis:02}. The notion of $n$-adic concept lattice can be introduced in the similar way to the triadic case \cite{Voutsadakis:02}. For mining $n$-adic formal concepts one can use \textsc{Data-Peeler} algortihm described in \cite{Cerf:2009}.

\subsection{Biclustering}

An alternative approach to define patterns in formal contexts can be realised via a relaxation of the definition of formal concept as a maximal rectangle full of crosses w.r.t the input incidence relation. One of such relaxations is the notion of an object-attribute bicluster \cite{Ignatov:2010}.
If $(g, m)\in I$, then $(m', g')$ is called an \emph{object-attribute bicluster}\footnote{we omit curly brackets here it what follows implying that $\{g\}'=g'$ and  $\{m\}'=m'$} (OA-bicluster or simply bicluster if there is no collision) with the density $\rho(m',g')=|I\cap(m'\times g')|/(|m'|\cdot|g'|)$.

\begin{figure}[ht]
  \centering
    \includegraphics[scale=0.45]{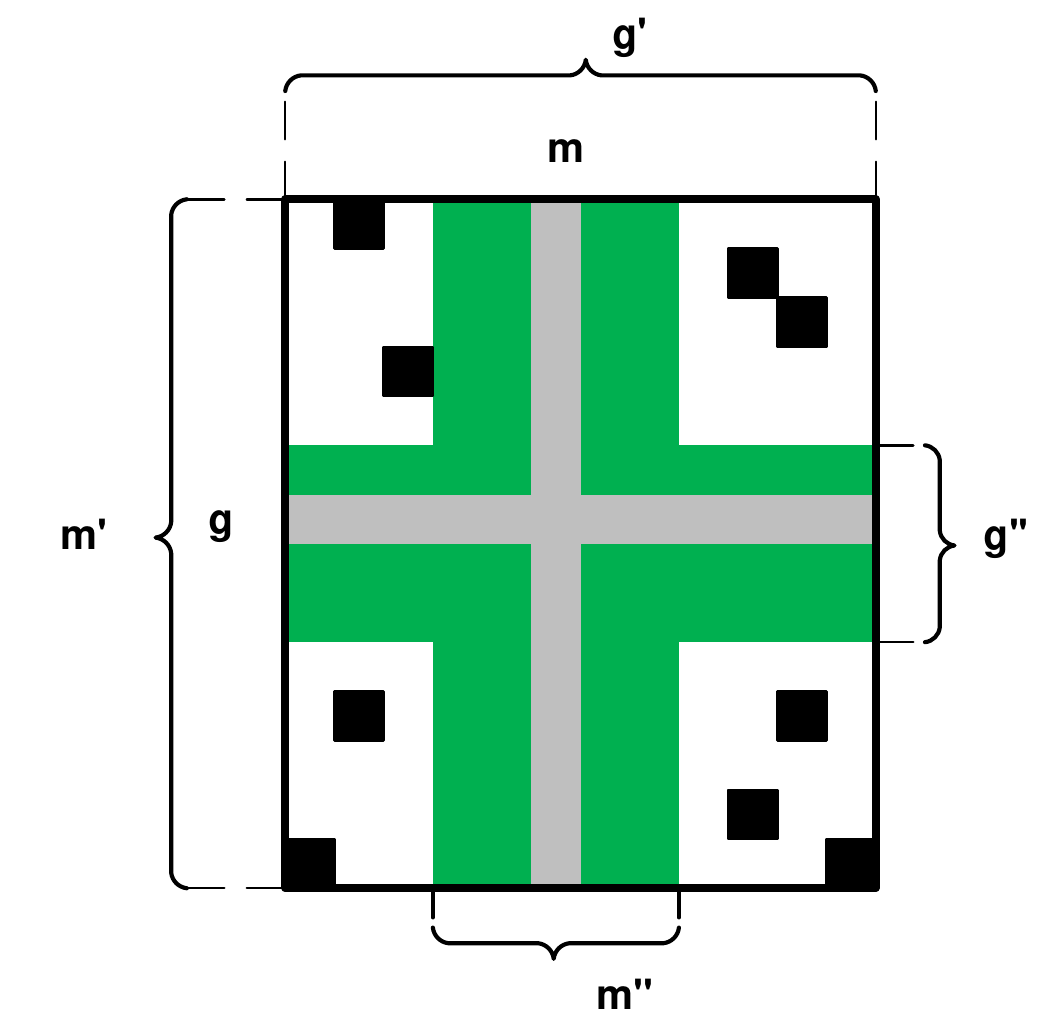}\\

  \caption{OA-bicluster.}\label{bicl}
\end{figure}

The main features of OA-biclusters are listed below:
\begin{enumerate}
 \item For any bicluster $(m',g')\subseteq 2^G\times 2^M$ it follows that $\frac{|m'|+|g'|-1}{|g'||m'|} \leq \rho(A,B)\leq 1$.
 \item OA-bicluster $(m',g')$ is a formal concept iff $\rho=1$.
 \item If $(m', g')$ is a bicluster, then $(g'', g')\leq (m', m'')$.
\end{enumerate}

Let $(A,B)\subseteq 2^G \times 2^M$ be a bicluster and $\rho_{min}$ be a non-negative real number such that $0 \leq\rho_{\min}\leq 1$, then $(A, B)$ is called \emph{dense}, if it fits the constraint $\rho (A,B) \geq \rho_{\min}$.
The above mentioned properties show that OA-biclusters differ from formal concepts by the fact that they do not necessarily have unit density. Graphically it means that not all the cells of a bicluster must be filled by a cross (see Fig.~\ref{bicl}). The rectangle in figure \ref{bicl} depicts a bicluster extracted from an object-attribute table. The horizontal gray line corresponds to object $g$ and contains only non-empty cells. The vertical gray line corresponds to attribute $m$ and also contains only non-empty cells. By applying the Galois operator, as explained in section \ref{FCA}, one time to $g$ we obtain all its attributes $g'$. By applying Galois operator $'$ twice to $g$ we obtain all objects that have the same attributes as $g$. This is depicted in Fig.~\ref{bicl} as $g''$. By applying Galois operator $'$ twice to $m$ we obtain all attributes that belong to the same objects as $m$. This is depicted in Fig.~\ref{bicl} as $m''$. The white spaces indicate empty cells. The filled black boxes indicate non-empty cells. Whereas a traditional formal concept would cover only the green and gray area, the bicluster also covers the white and black cells. This gives to OA-biclusters fault-tolerance properties (see Proposition~\ref{tric_save_prop}).

%\begin{algorithm}
%\caption{Bicluster computation\label{BiA}}
%\DontPrintSemicolon
%\LinesNumbered
%\KwData{$K=(G,M,I)$ is a formal context, $\rho_{min}$ is a threshold density value of bicluster density}
%\KwResult{$B=\{(A_k,B_k)| (A_k,B_k)$ -- bicluster$\}$ }
%\Begin{
%$Obj.Size=|G|$\;
%$Attr.Size=|M|$\;
%$B \longleftarrow \emptyset$\;
%\For{$g\in G$}{
%$Obj[g]=g'$\;
%}
%\For{$m \in M$}{
%$Attr[m]=m'$\;
%}
%\For{$g \leftarrow 0$ \KwTo $|G| $}{
%\For{$m \in Obj[g]$}{
%\If{$\rho(Attr[m],Obj[g])\geq \rho_{min}$}{$B.Add((Attr[m],Obj[g]))$}
%}
%}
%}
%\end{algorithm}

\begin{algorithm}
\caption{Add procedure for the online algorithm for OA-biclustering.}
\label{BiA}
    \begin{algorithmic}[1]
    	\REQUIRE $I$~is an input set of object-attribute pairs;\\
            $\mathcal{B}=\{B=(*X,*Y)\}$~is a current set of OA-biclusters;\\
            $PrimesOA$, $PrimesAO$;\\
    	\ENSURE $\mathcal{B}=\{T=(*X,*Y)\}$;\\
            $PrimesOA$, $PrimesAO$;
    	\FORALL{$(g,m)\in I$}
    		\STATE $PrimesOA[g]:=PrimesOA[g]\cup m$
            \STATE $PrimesAO[m]:=PrimesAO[m]\cup g$
           
            \STATE $\mathcal{B}:=\mathcal{B}\cup (\&PrimesAO[m],\&PrimesOA[g])$
    	\ENDFOR
    \end{algorithmic}
\end{algorithm}

To generate biclusters fulfilling a minimal density requirement we can  perform computations in two phases. The online phase, Add procedure (see Algorithm~\ref{BiA}), allows to process pairs from incidence relation $I$ and generate biclusters in one pass  by means of pointer and reference variables for access to primes of objects and attributes even without knowing the number of objects and attributes in advance; see the version of this online algorithm for triadic case in \cite{Gnatyshak:2014}. Thus, generation of all biclusters is realised within $O(|I|)$. Note that the algorithm can start with a non-empty collection of biclusters obtained previously. Then all biclusters can be enumerated in a sequential manner and only those fulfilling the minimal density constraint are retained.

\begin{svgraybox}
For the context shown in Fig.~\ref{fig:readers} one can find two concepts,

 $C_2=(\{Alex, David\}, \{The \ Puppet \ Masters, Ubik\})$ and 

 $C_4=(\{Alex, Mike, David\}, \{Ubik\}),$ and one bicluster,

 $B_1=(Ubik',David')=(\{Alex, Mike, David\}, \{The \ Puppet \ Masters, Ubik\})$,  with density $\rho=5/6\approx0.83$. 
 
These two concepts can be interpreted as Sci-Fi readers and cyber punk readers (or P.K. Dick's readers at least), respectively. However, bicluster $B_1$ by allowing one missing pair $(Mike, The \ Puppet \ Masters)$ can be considered as a community of Sci-Fi readers as well, which is larger than $C_2$.

\end{svgraybox}

\subsection{OAC-Triclustering and Prime-based $n$-clustering}
Guided by the idea of finding scalable and noise-tolerant alternatives to triconcepts, we have had a look at triclustering paradigm in general for a triadic binary data, i.e. for tricontexts as input datasets.

\begin{definition}
    Suppose $\mathbb{K}=(G,M,B,I)$ is a triadic context and $Z \subseteq G$, $Y \subseteq M$, $Z \subseteq B$.
    A triple $T=(X,Y,Z)$ is called an \emph{OAC-tricluster}.
    Traditionally, its components are respectively called \emph{extent, intent, and modus}.
\end{definition}

The \emph{density} of a tricluster $T=(X,Y,Z)$ is defined as the fraction of all triples of $I$ in $X\times Y\times Z$:

\begin{equation}
    \rho(T)=\frac{|I\cap(X\times Y\times Z)|}{|X||Y||Z|}
\end{equation}

\begin{definition}
    A tricluster $T$ is called \emph{dense} iff its density is not less than some predefined threshold, i.e. $\rho(T)\ge\rho_{min}$.
\end{definition}

The collection of all triclusters for a given tricontext $\K$ is denoted by $\mathcal{T}$.

Since we deal with all possible cuboids in Cartesian product $G\times M\times B$, it is evident that the number of all OAC-triclusters, $|\mathcal{T}|$, is equal to $2^{|G|+ |M|+ |B|}$. However not all of them are supposed to be dense, especially for real data which are frequently quite sparse. Below we discuss one of possible OAC-tricluster definitions, which give us an efficient way to find, within polynomial time, a number of (dense) triclusters not greater than the number of triples in the initial data, $|I|$.

Here, let us define the prime operators and describe \emph{prime OAC-triclustering}, which extends the biclustering method from \cite{Ignatov:2012a} to the triadic case.

Derivation (prime) operators for elements of a triple $(\widetilde{g},\widetilde{m},\widetilde{b})\in I$ from a triadic context $\mathbb{K}$ can be defined as follows:

\begin{equation}
	\widetilde{g}^\prime:=\{\,(m,b)\mid(\widetilde{g},m,b)\in I\} \label{prime-o}
\end{equation}
\begin{equation}
	\widetilde{m}^\prime:=\{\,(g,b)\mid(g,\widetilde{m},b)\in I\} \label{prime-a}
\end{equation}
\begin{equation}
	\widetilde{b}^\prime:=\{\,(g,m)\mid(g,m,\widetilde{b})\in I\} \label{prime-c}
\end{equation}

$(\widetilde{g},\widetilde{m})^\prime$, $(\widetilde{g},\widetilde{b})^\prime$, $(\widetilde{m},\widetilde{b})^\prime$ prime operators can be defined in the same way.

\begin{equation}
	(\widetilde{g},\widetilde{m})^\prime:=\{\,b\mid(\widetilde{g},\widetilde{m},b)\in I\} \label{prime-oa}
\end{equation}
\begin{equation}
	(\widetilde{g},\widetilde{b})^\prime:=\{\,m\mid(\widetilde{g},m,\widetilde{b})\in I\} \label{prime-oc}
\end{equation}
\begin{equation}
	(\widetilde{m},\widetilde{b})^\prime:=\{\,g\mid(g,\widetilde{m},\widetilde{b})\in I\} \label{prime-ac}
\end{equation}

The following definition uses only prime operators (eqs. \ref{prime-oa}--\ref{prime-ac}) to generate triclusters, however other variants are possible. Thus, in~\cite{Ignatov:2015}, \emph{box operator based OAC-triclusters} have been studied; this type of tricluster relies on \ref{prime-o}--\ref{prime-c}.

\begin{definition}
    Suppose $\mathbb{K}=(G,M,B,I)$ is a triadic context.
    For a triple $(g,m,b)\in I$ a triple $T=\left((m,b)^\prime,(g,b)^\prime,(g,m)^\prime\right)$ is called a \emph{prime operator based OAC-tricluster}.
    Its components are called respectively \emph{extent, intent, and modus}.
\end{definition}

\begin{figure}[t]
\begin{center}
	\includegraphics[scale=0.7]{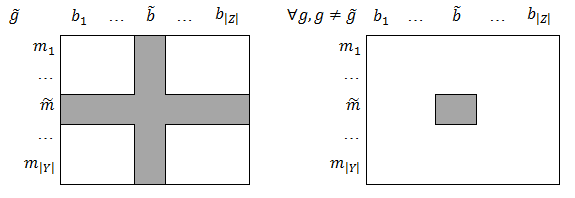}
	\caption{Prime operator based tricluster structure}
	\label{prime-struct}
\end{center}
\end{figure}

Prime based OAC-triclusters are more dense than box operator based ones.
Their structure is illustrated in Fig.~\ref{prime-struct}: every element corresponding to the ``grey'' cell is an element of $I$.
Thus, prime operator based OAC-triclusters in a three-dimensional matrix (tensor) form contain an absolutely dense cross-like structure of crosses (or ones).

The proposed OAC-tricluster definition has a fruitful property (see Proposition~\ref{tric_save_prop}): for every triconcept in a given tricontext there exists a tricluster of the same tricontext in which the triconcept is contained w.r.t. component-wise inclusion. It means that there is no information loss, we keep all the triconcepts in the resulting tricluster collection.

\begin{proposition}
\label{tric_save_prop} Let $\mathbb{K}=(G,M,B,I)$ be a triadic context and $\rho_{min}=0$. For every
$T_c=(X_c,Y_c,Z_c) \in \mathfrak{T}(G,M,B,I)$ with non-empty $X_c$, $Y_c$, and $Z_c$  there exists a prime OAC-tricluster $T=(X,Y,Z) \in \mathcal{T}_{\prime}(G,M,B,Y)$ such that $X_c \subseteq X,Y_c \subseteq Y, Z_c \subseteq Z$.
\end{proposition}

\noindent (Here, $\mathcal{T}_{\prime}(G,M,B,I)$ denotes the set of all OAC-prime tricluters fulfilling the chosen value of $\rho_{min}$.)

\begin{proof}
 Let $(g,m,b) \in X_c \times Y_c \times Z_c$. By the definition of prime operators $(m,b)^\prime:=\{\,\widetilde{g} \mid(\widetilde{g},m,b)\in I\}$.  Since $m \in Y_c$ and $b \in Z_c$ then by the definition of formal triconcept $(m,b)$  is related by $Y$ to every $\tilde{g} \in X_c$, therefore $(m,b)^\prime \cap X_c =X_c $. Consequently for all $g_i \in X_c$ we have $g_i \in (m,b)^\prime$. For  $(g,b)^\prime$ and $(g,m)^\prime$ tricluster components the proof is similar. Finally, we have  $X_c \subseteq X=(m,b)^\prime, Y_c \subseteq Y=(g,b)^\prime, \mbox{ and } Z_c \subseteq Z=(g,m)^\prime$.
\end{proof}

Prime-based $n$-clustering can be introduced similarly. Let $\K=(X_1,X_2, \ldots, X_n, Y)$ be an $n$-adic context and $Y$ is binary relation between $X_1 \ldots X_n$.

Then for a tuple $(x_1,x_2,\ldots,x_n) \in Y$ we define $n$ prime operators for each tuple $(x_1,\ldots,x_{i-1},x_{i+1},\ldots,x_n)$ as follows:

\[(\{x_1\},\ldots,\{x_{i-1}\},x_{i+1},\ldots,\{x_n\})'=\{ z_i \mid  (x_1,\ldots,x_{i-1},z_{i},x_{i+1},\ldots,x_n) \in Y \}.\]

For a given tuple $(x_1,x_2,\ldots,x_n) \in Y$, a prime operator based $n$-cluster $P$ is defined as follows:

$$P=((\{x_{2}\},\ldots,\{x_n\})',\ldots, (\{x_1\},\ldots,\{x_{i-1}\},\{x_{i+1}\},\ldots,\{x_n\}\})', \ldots,$$
$$ (\{x_1\},\ldots,\{x_{n-1}\})').$$

The density of $n$-cluster $P=(Z_1,Z_2,\ldots,Z_n)$ is $\rho(P)=\frac{|Y \cap Z_1 \times Z_2 \times \ldots \times Z_n|}{|Z_1 \times Z_2 \times \ldots \times Z_n|}$. To keep analogy of $\rho$ with physical density we refer to its enumerator as the mass of $P$, i.e. $mass(P)$, while its denominator plays a role of the volume of $P$, i.e. $vol(P)$.

The description of a one-pass algorithm for OAC-prime tricluster generation can be found in \cite{Gnatyshak:2014}. A Map-Reduce based prototype of OAC-prime triclustering and possible implementation variants are presented in \cite{Zudin:2015}.

\section{Quality measures for multimodal clustering}\label{sec:quality}

\subsection{Connection between $\rho$ and local clustering coefficient}

Since we use density as a local measure of $n$-cluster quality, it is useful to find its connection to local clustering coefficients (we use $cc_{\bullet}(\cdot)$ notation from \cite{Latapy:2008}).
For $(V,E \subseteq V\times V)$, the local clustering coefficient is $cc_{\bullet}(v)=\frac{|N(v) \times N(v) \cap E|}{N(v)(N(v)-1)/2}$, here $N(v)$ is the degree of $v \in V$.

If one considers a 1-mode network $(V,E \subseteq V\times V)$ as a formal context $\K=(G,G,I\subseteq G \times G)$, where $V=G$, and for $g,m \in V$ $gEm \iff gIm$,  then for bicluster $(g',g')$ it follows that\footnote{Note that technically $(g',g')$ is not an OA-bicluster since $(g,g) \not \in I$}

$$\rho(g',g')=\frac{|g'\times g' \cap I|}{|g'||g'|}=\frac{|N(g)\times N(g)\cap I|}{|N^2(g)|}=\frac{|N(g)\times N(g)\cap I|}{\frac{(|N(g)|-1)|N(g)|}{2}}\frac{1-1/|N(g)|}{2}=$$
$$=cc_{\bullet}(g)\frac{1-\frac{1}{|N(g)|}}{2}.$$

Note that $N(g)=deg(g)=\{u|  gEu\}=g'$.

Moreover, for large neighbourhoods $\rho(g',g')\approx \frac{cc_{\bullet}(g)}{2}$.

\subsection{Connection between $\rho$ and modularity}

Since we do not optimise any modularity-like criterion in our study, multimodal clusters supposed to be overlapped in general, and, moreover, to the best of our knowledge there is no widely accepted modularity criterion even for bipartite overlapped communities; the introduction and study of such criteria could be a subject of a separate research. However, we show the interconnection between average sum of values in the input modularity matrix for a particular bicluster and its density.

Let $A_{gm}$ be the adjacency matrix of an input context $\K=(G,M,I\subseteq G \times M)$, i.e. 
$A_{gm}=[gIm]$\footnote{Here $[\cdot]$ means Iverson bracket defined as $[P] = \begin{cases} 1 & \text{if } P \text{ is true;} \\ 0 & \text{otherwise,} \end{cases}$} for $(g,m)\in G \times M$. For bipartite graphs an entry of modularity matrix is defined as follows:

$$B_{gm}=A_{gm}-\frac{deg(g)deg(m)}{|I|}=[gIm]-\frac{|g'||m'|}{|I|} .$$

For non-overlapped communities modularity in two-mode networks is defined as follows \cite{Barber:2007}:

$$Mod=\frac{1}{|I|}\sum\limits_{(g,m) \in G \times M}\left([gIm]-\frac{|g'||m'|}{|I|}\right)[(g,m) \in C], \mbox {where}$$

$C \subseteq G \times M$ is a module (or community) from a set of non-overlapped communities $\mathcal{C}$ of the original network. Non-overlapping here is formally defined as follows: $\forall C, D \in \mathcal{C}$ $C \cap D = \emptyset$.

Let $(m',g')$ be a bicluster of $\K$, then the sum over all entries $(\tilde{g}, \tilde{m}) \in m' \times g'$ in $B$ gives: 

$$|m' \times g' \cap I| - \frac{\sum\limits_{(\tilde{g}, \tilde{m}) \in m' \times g'}|\tilde{g}'||\tilde{m}'|}{|I|}. $$

Instead of normalising that sum by $|I|$ as in modularity definition, we can try to calculate (local) bicluster modularity, $Mod_l(m',g')$, by normalising the sum by the bicluster volume $Vol(m',g')=|g'||m'|$:

$$Mod_l(m',g')=\frac{|m' \times g' \cap I|}{|g'||m'|}- \frac{\sum\limits_{\tilde{g} \in m'}|\tilde{g}'|\sum\limits_{ \tilde{m} \in g'}|\tilde{m}'|}{|g'||m'||I|}=\rho(m',g')-\frac{\overline{deg}(\tilde{g})\overline{deg}(\tilde{m})}{|I|}, \mbox{ where}$$

$\overline{deg}(\tilde{g})=\frac{\sum_{\tilde{g} \in m'}}{|g'|}$ is the average degree of $\tilde{g}$ in the input bicluster and $\overline{deg}(\tilde{m})$ is the average degree of $\tilde{m}$ and defined similarly.

It is clear, that to maximise $Mod_l$ criterion one need to find a bicluster with high density and low average degrees of its elements.

However, the original modularity criterion for bipartite non-overlapped networks has intrinsic drawbacks.
One of them is low resolution problem lying in dependence between the size of detected communities and the size of an input graph \cite{Fortunato:2010}. Another one can be demonstrated by a model example.

\begin{svgraybox}
 Let $\K=(G,M, I)$ be a formal context, where for a certain pair $(g,m) \in I$ we have $g'=M, m'=G$, and $I=m' \times g'$. Without loss of generality let $|G|=|M|=n$. Then 

$$B_{gm}=[gIm]-\frac{|g'||m'|}{|I|}=1-\frac{n^2}{2n-1} .$$

For large $n$, $B_{gm} \approx 1-n/2$ and this value tends to $-\infty$ by implying $n \to \infty$. To keep the second term of an entry of the modularity matrix no greater than 1 (the maximal probability of incidence of $g$ and $m$), one needs to require $|g'|, |m'| \leq \sqrt{|I|}$ (which is in fact should be normally fulfilled for large and sparse (real) networks).
\end{svgraybox}

\subsection{Least square optimal $n$-clusters}

One of the important statistics in Clustering is the data scatter of an input matrix, i.e. the sum of squares of all its entries \cite{Mirkin:1996}. In \cite{Mirkin:2011}, lest squares based maximisation criterion to generate $n$-cluster was proposed:

$$g(P)=\rho^2(P) \cdot Vol(P)= \rho(P) \cdot mass(P), \mbox{ where}$$

$P$ is an $n$-cluster of a certain $n$-adic context. On the one hand, its direct interpretation implies that we care about dense $n$-clusters of large size instead of only dense (that may be small) or only large (that may be sparse); in other words such $n$-clusters tend to be massive (with low number of missing tuples in the input binary relation) and dense. On the other hand,  this criterion measures the contribution of $P$ to the data scatter of the input $n$-adic context.

In \cite{Ignatov:2015}, one can find a theorem saying that by maximisation of $g(P)$ we require higher density within $n$ cluster $P$ than in the corresponding outside regions along its dimensions.

\subsection{Weak bicluster communities and graph cuts}

In network analysis, a community is called weak if its average internal degree is greater than its average out degree \cite{Barabasi:2016}.

In two-mode case, for an input context $\K=(G,M,I)$ and its bicluster $(m',g')$, we have:

$$\sum\limits_{\tilde{g}\in m'} |(\{\tilde{g}\} \cup \{g\})'| +  \sum\limits_{\tilde{m}\in g'}|(\{\tilde{m}\} \cup \{m\})'| \geq \sum\limits_{\tilde{g}\in m'} |\tilde{g}' \cap M \setminus g' | +  \sum\limits_{\tilde{m}\in g'} |\tilde{m}' \cap G \setminus m' |.  $$

The left handside of the inequality is the doubled sum of the number of object-attribute pairs from $(m',g')$. The right handside shows how many pairs object from bicluster extent and attributes from bicluster input form with remaining attributes and objects of the context. In network analysis this measure is known as $cut$ \cite{Fortunato:2010}, i.e. the number of edges one should delete to make the community disconnected from the remaining vertices in the input graph. Thus, the inequality can be rewritten as follows:

 $$\rho(m',g') \geq \frac{cut(m',g')}{2|g'||m'|}.$$ 

This criterion can be used for selection of biclusters during their generation instead of fixed $\rho_{min}$.

\subsection{Stability of OA-biclusters}

\emph{Stability of formal concepts}~\cite{Kuznetsov:1990,Kuznetsov:2007} has been used as a means of concepts' filtering in studies on epistemic communities \cite{Roth:2006,Kuznetsov:2007a,Roth:2008} and communities of website visitors c\cite{Kuznetsov:2007w}.

Let $\K = (G, M, I)$ be a formal context and $(A, B)$ be a formal concept of $\K$. The \emph{(intensional) stability index}, $\sigma$, of $(A,B)$ is defined as follows:
$$\sigma(A,B) = \frac{|\{C \subseteq A \mid  C' = B\}|}{ 2ˆ{|A|}}$$

As we know, not all of the OA-buclusters of a given formal context are formal concepts. 

Only those OA-biclusters that fulfil condition $(m',g')=(g'',m'')$ are formal concepts. However, stability index can be technically computed for any OA-bicluster as follows:

$$\sigma(m',g') = \frac{|\{A \subseteq m' \mid A' = g'\}|}{ 2^{|m'|}}$$

Set $2^{m'}$ can be decomposed into three parts:  $2^{g''} \cup 2^{m'\setminus g''}\cup \Delta$.
The enumerator is equal to $|\{A \in 2^{g''} \mid A' = g'\}|+|\{ A \in 2^{m'\setminus g''} \mid A' = g'\}\setminus \emptyset|+|\{ A \in \Delta \mid A' = g'\}\setminus \emptyset|$. Since every set of objects from $m'\setminus g''$ does not have all attributes from $g'$, the second summand is 0, and the same applies to the third one due to each set from $\Delta$ contains at least one object $\tilde{g}$ from $m'\setminus g''$ such that $\tilde{g'}\not = g'$. Hence, 

$$\sigma(m',g') = \frac{|\{A \in 2^{g''} \mid A' = g'\}|}{ 2^{|m'|}}.$$

Since the number of all $A$ that contain $g$ is $|2^{g''\setminus g}|$, the tight lower bound of OA-bicluster's stability is $2^{|{g''\setminus g}|-|m'|}$.

The stability index of a concept indicates how much the concept intent depends on particular objects of the extent.

\subsection{Coverage and diversity}

\emph{Diversity} is an important measure in Information Retrieval for diversified search results and in Machine Learning for ensemble construction \cite{Tsymbal:2005}.

To define diversity for multimodal clusters we use a binary function  that equals to 1 
if the intersection of triclusters ${T}_i$ and ${T}_j$ is not empty, and 0 otherwise.

\begin{equation}
\label{intersect}
	intersect({T}_i,{T}_j)=
	\left[ G_{{T}_i}\cap G_{{T}_j}\not=\emptyset\wedge M_{{T}_i}\cap M_{{T}_j}\not=\emptyset\wedge B_{{T}_i}\cap B_{{T}_j}\not=\emptyset \right]
\end{equation}

It is also possible to define $intersect$ for the sets of objects, attributes and conditions. For instance, $intersect_G({T}_i,{T}_j)$ is equal to 1 if triclusters ${T}_i$ and ${T}_j$ have nonempty intersection of their extents, and 0 otherwise.

Now we can define \emph{diversity of the tricluster set} $\mathcal{T}$:

\begin{equation}
\label{diversity}
	diversity(\mathcal{T})=1-\frac{\sum_j\sum_{i<j}intersect({T}_i,{T}_j)}{\frac{|\mathcal{T}|(|\mathcal{T}|-1)}2}
\end{equation}

The \emph{diversity for the sets of objects (attributes or conditions)} is similarly defined:

\begin{equation}
\label{diversity2}
	diversity_G({T})=1-\frac{\sum_j\sum_{i<j}intersect_G({T}_i,{T}_j)}{\frac{|\mathcal{T}|(|\mathcal{T}|-1)}2}
\end{equation}

\noindent \textbf{Coverage} is defined as a fraction of the triples of the context (alternatively, objects, attributes or conditions) included in at least one of the triclusters of the resulting set.

More formally, let $\K=(G,M,B,I)$ be a tricontext and $\mathcal{T}$ be the associated triclustering set obtained by some triclustering method, then coverage of $\mathcal{T}$:

\begin{equation}
\label{coverage}
coverage(\mathcal{T})=\sum\limits_{(g,m,b)\in I} \left[(g,m,b)\in \bigcup\limits_{(X,Y,Z)\in \mathcal{T}} X \times Y \times Z\right]/|I|.
\end{equation}

The \emph{coverage of the object set} $G$ by the tricluster collection $\mathcal{T}$ is defined as follows:

\begin{equation}
\label{o_coverage}
coverage_G(\mathcal{T})=\sum\limits_{g\in G} \left[g\in \bigcup\limits_{(X,Y,Z)\in \mathcal{T}} X \right]/|G|.
\end{equation}

\emph{Coverage of attribute or condition sets} can be defined analogously. These measures may have sense when would like to know how many actors or items in the network do not belong to any found community.

We also use the \emph{coverage of formal concepts by biclusters}, i.e.  we count the number of concepts covered by at least one bicluster in the corresponding bicluster collection $B$. We say that bicluster  $B=(X,Y)$ covers concept $C=(Z,W)$ w.r.t.  component-wise inclusion of their extents and intents, namely $C \sqsubseteq B :\iff  Z \subseteq X \mbox{ and } W \subseteq Y $.

\begin{equation}
\label{bic_coverage}
coverage_\mathcal{B}(\BGMI )=  \frac{\{ C \in \BGMI  \mid  \exists B \in \mathcal{B}: C \sqsubseteq B \}}{|\BGMI|}. 
\end{equation}

\section{Data}\label{data}

For our experiments we collected datasets from 1-mode to 4-mode networks.

In particular, we have analysed the following classic 1-mode datasets:

\begin{itemize}

\item Karate club,	34$\times$34, 78 edges;	
\item Florent family 1,		16$\times$16, 40 edges;
\item  Florent family 2,		16$\times$16, 30 edges;
\item Hi-tech, 36$\times$36, 147 edges;
\item Mexican people, 35$\times$35, 117 edges.

\end{itemize}

For 2-mode datasets we have used Southern women of size 18x14 with 93 edges and four datasets studied in \cite{Latapy:2008}:

\begin{itemize}

\item co-authoring, 19,885$\times$16,400, and 45,904 edges;

\item co-occurrence, 13,587$\times$9,263, and 1,833,63 edges;

\item actor, 127,823$\times$383,640, and 1,470,418 edges;

\item p2p, 1,986,588 peers$\times$5,380,546 data, and  55,829,392 links (edges);

\end{itemize}

As for three-mode network, we have analysed Bibsonomy dataset \footnote{\url{http://www.kde.cs.uni-kassel.de/bibsonomy/dumps/}} with 
$|U| =$ 2,467 users, $|T| =$ 69,904 tags, $|R| =$ 268,692 resources that related by $|Y| =$ 816,197 triples.

Finally, MovieLens data\footnote{\url{http://grouplens.org/datasets/movielens/}} with 100,000 ratings (integers from 1 to 5) and 1,300 tag applications applied to 9,000 movies by 700 users is considered as a 4-mode dataset. We have used only user, movie, rating and time modes.

%For our experiments we collected a dataset from the Russian social networking site Vkontakte. Each entry consisted of the following fields: id, userid, gender, family status, birthdate, country, city, institute, interests, groups. This set was divided into 4 subsets based on the values of the institute field, namely students of two major technical universities and two universities focusing on humanities and sociology were considered: The Bauman Moscow State Technical University, Moscow Institute of Physics and Technology (MIPT), the Russian State University for Humanities (RSUH) and the Russian State Social University (RSSU). Then 2 formal contexts, users-interests and users-groups were created for each of these new subsets.
%	
%\begin{table}[ht]
%\caption{Basic description of four data sets of large Russian universities.}\label{tbl-univ}
%\begin{center}
%\begin{tabular}{c|cccc}
%  \hline
%  % after \\: \hline or \cline{col1-col2} \cline{col3-col4} ...
%   & Bauman &	MIPT	& RSUH &	RSSU\\
%  \hline
%
%number of users &	18542 &	4786 &	10266 &	12281\\
%number of interests &	8118 &	2593 &	5892 & 3733\\
%number of groups &	153985 &	46312 &	95619 &	102046\\
%  \hline
%\end{tabular}
%
%\end{center}
%\end{table}

\section{Experiments}\label{exp}

We have tested our implementations for one- and two-mode networks in Python 2.7 and for higher modes in C\# with our tool, Multimodal Clustering Toolbox, on a Mac Pro computer with 3.7 GHz and 16 GB RAM.

\subsection{Two-mode networks}

For each two-mode dataset we report the number of unique biclusters and the number of all generated  biclusters; note that when all objects (and attributes) are pairwise different there are no duplicates by definition.

For small and medium size classic two-mode and one-mode datasets we have reported the number of formal concepts covered by the generation bicluster collection for a specific $\rho_{min}$ as well as their fraction, i.e. $coverage_\mathcal{B}(\BGMI)$.

In 1930s, a group of ethnographers collected data on the social activities of 18 women over a nine-month period~\cite{Davis:1941}. Different subgroups of these women had met in 14 informal social events; the incidence of a woman to a particular event was established using ``interviews, the records of participant observers, guest lists, and the newspapers'' (\cite{Davis:1941}, p. 149). Later on, this Souther Women data set has become a benchmark for comparing communities detection methods in two-mode social network analysis, in particular, including concept lattices as a community detection approach~\cite{Freeman:1993,Freeman:2003}\footnote{There is a small inconsistency in the profiles of women $w_{14}$ (Helen) and $w_{15}$ (Dorothy), namely  between their description in \cite{Freeman:2003} and the downloaded dataset provided at \url{https://networkdata.ics.uci.edu/netdata/html/davis.html}, thus according to the latter $e_{12}, e_{13} \in w_{14}'$ and $e_{11}, e_9 \in w_{15}'$}. 

\begin{table}[t!]
\caption{Southern women: 18x14, 93 edges}
\centering
\begin{tabular}{c|c|c|c|c}
$\rho$	& concept  &	Unique  & biclusters	& Fraction of \\
	&  coverage &	biclusters & biclusters	& covered concepts\\
\hline 
0 & 65 & 83 & 93 & 1.00\\
0.05 & 65 & 83 & 93 & 1.00\\
0.1 & 65 & 83 & 93 & 1.00\\
0.15 & 65 & 83 & 93 & 1.00\\
0.2 & 65 & 83 & 93 & 1.00\\
0.25 & 65 & 83 & 93 & 1.00\\
0.3 & 65 & 83 & 93 & 1.00\\
0.35 & 65 & 82 & 92 & 1.00\\
0.4 & 65 & 81 & 91 & 1.00\\
0.45 & 65 & 77 & 87 & 1.00\\
0.5 & 65 & 71 & 81 & 1.00\\
0.55 & 65 & 63 & 73 & 1.00\\
\hline
0.6 & 65 & 60 & 7 & 1.00\\
0.65 & 64 & 51 & 59 & 0.98\\
\hline
0.7 & 63 & 40 & 47 & 0.97\\
0.75 & 57 & 33 & 4 & 0.88\\
0.8 & 51 & 22 & 28 & 0.78\\
0.85 & 35 & 13 & 19 & 0.54\\
0.9 & 20 & 7 & 9 & 0.31\\
0.95 & 0 & 0 & 0 & 0.00\\
1 & 0 & 0 & 0 & 0.00\\
\hline 
\end{tabular} 
\end{table}

\begin{figure}
\centering
\includegraphics[width=0.95\textwidth]{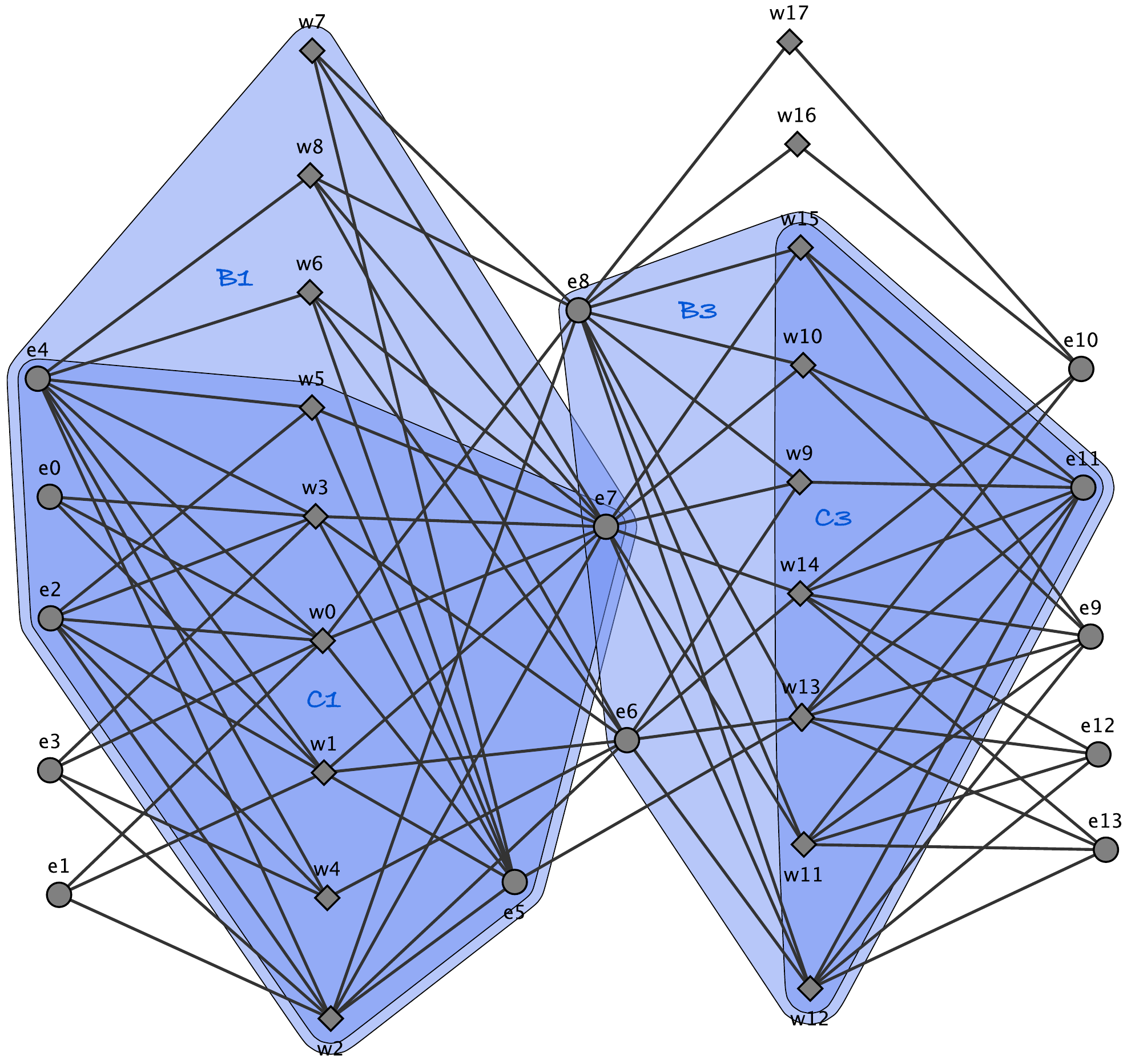}
\caption{The two-mode network for the Southern women dataset, bicluster $B_1$ and concept $C_1$, and bicluster $B_3$ and concept $C_3$}\label{fig:women}
\end{figure}

\begin{svgraybox}
There are 66 formal concepts for the Southern woman network. Since OA-biclusters are tolerant to missing values, let us illustrate how rather dense biclusters include the largest concepts with non-empty extent and intent. 

For example, with $\rho_{min}=0.8$ we show five bicluster-concept pairs $B_i=(e',w')$, $C_i=(W,E)$ related by component-wise inclusion of their extents and intents, respectively, namely $C_i \sqsubseteq B_i :\iff  W \subseteq e' \mbox{ and } E \subseteq w' $ : 

\be

\item 	
$C_1=(\{ w_0, w_1, w_2, w_3, w_5, w_6, w_7\},\{e_5, e_7\}) \sqsubseteq B_1=(\{w_0, w_1, w_2, w_3, w_5, w_6, w_7, w_8\},\{e_2, e_4, e_5, e_7\})$  with $\rho(B_1)=0.84$;

\item 
%0.82		
%0,2,3,4		0,2,3,4,5,6,7
%U
%0,2,3		2,3,4,5,7

$C_2=(\{ w_0, w_2, w_3\},\{e_2, e_3, e_4, e_5, e_7\}) \sqsubseteq B_2=(\{w_0, w_2, w_3, w_4\},\{e_0, e_2, e_3, e_4, e_5, e_6, e_7\})$  with $\rho(B_2)=0.82$;

\item 
%0.82		
%9,10,11,12,13,14,15		8,11,6,7
%	U	
%9,10,11,12,13,14,15		11

$C_3=(\{ w_9, w_{10}, w_{11}, w_{12}, w_{13}, w_{14}, w_{15}\},\{e_{11}\}) \sqsubseteq B_3=(\{ w_9, w_{10}, w_{11}, w_{12}, w_{13}, w_{14}, w_{15}\},\{e_6, e_7, e_8, e_{11}\})$  with $\rho(B_3)=0.82$;

\item
%0.92		
%10,11,12,13,14,15		8,9,11,7
%55	U	
%10,11,12,15		8,9,11,7
%53			

$C_4=(\{ w_{10}, w_{11}, w_{12}, w_{15}\},\{e_7, e_8, e_9, e_{11}\}) \sqsubseteq B_4=(\{w_{10}, w_{11}, w_{12}, w_{13}, w_{14}, w_{15}\},\{e_7, e_8, e_9, e_{11}\})$  with $\rho(B_4)=0.92$;	
		
\item 
%0.88 
%16,17,13,14		8,1
%	U	
%16,17,13		8,1

$C_5=(\{ w_{16}, w_{17}, w_{13}\},\{e_1, e_8\}) \sqsubseteq B_5=(\{w_{16}, w_{17}, w_{13}, w_{14}\},\{e_1,e_8\})$  with $\rho(B_5)=0.88$.

\ee

The corresponding bipartite graph is shown in Fig.~\ref{fig:women} along with the biclique formed by elements of concept $C_1$ and bicluster $B_1$, and concept $C_3$ and bicluster $B_3$. According to \cite{Freeman:2003,Doreian:2004} there is   the ``true structure'' of the Southern women network; namely, there are two groups of women $\{w_0, \ldots, w_8\}$ and 
$\{w_1, \ldots, w_{17}\}$. The first group of women participated in events $e_0$ through $e_4$, while the second group was not. The second group participated in events $e_3$ through $e_{13}$, while the first group was not. Both groups participated $e_6$, $e_7$, and $e_8$.

\end{svgraybox}

Since the  Southern women network is a well-studied case in SNA community and one of the first SNA datasets analysed by sociologists using concept lattices, an interested reader may refer to \cite{Freeman:1993,Freeman:2003} to find professional interpretation of several important communities of women found by means of formal concepts.

Even though that such networks as co-authoring, co-occurrence, actor,  and p2p 
 are two-mode and known to SNA community about a decade, even the number of concepts (maximal bicliques) for these datasets is not reported in the literature.

\begin{table}
\caption{The numbers of unique and all OA-biclusters for the four large two-mode networks}\label{tbl:two-mode}
\centering
\begin{tabular}{l|c|c|c|c|c|c|c|c|}
\cline{2-9}
& \multicolumn{8}{c|}{Datasets}\\
 \cline{2-9}
 & \multicolumn{2}{c|}{co-authoring} & \multicolumn{2}{|c|}{co-occurence} & \multicolumn{2}{|c|}{actor} & \multicolumn{2}{|c|}{p2p}\\
  \cline{2-9}
$\rho$  &       unique  & biclusters    &	        unique  & biclusters    &	        unique  & biclusters    &	       unique  	&	 biclusters    \\
         &      biclusters & &	                biclusters & &	        biclusters & &	      biclusters 	&	 \\
\hline 					
0 & 43,253 & 45,904&	 161,386 & 183,363&	 1,278,989 & 1,470,418&	 54,789,256 	&	 55,829,169\\
0.05 & 43,253 & 45,904&	 161,386 & 183,363&	 1,226,429 & 1,417,827&	 41,937,580 	&	 42,973,016\\
0.1 & 43,253 & 45,904&	 160,200 & 181,630&	 962,389 & 1,153,704&	 27,178,639 	&	 28,196,480\\
0.15 & 43,253 & 45,904&	 124,383 & 137,367&	 700,207 & 891,401&	 18,320,253 	&	 19,321,315\\
0.2 & 43,251 & 45,902&	 69,283 & 75,761&	 523,446 & 714,509&	 13,179,196 	&	 14,165,402\\
0.25 & 43,184 & 45,835&	 39,081 & 43,252&	 410,118 & 601,065&	 9,789,039 	&	 10,759,880\\
0.3 & 42,748 & 41,774&	 24,484 & 27,672&	 318,245 & 509,068&	 7,019,097 	&	 7,969,965\\
0.35 & 41,774 & 44,423&	 17,011 & 19,718&	 269,642 & 460,361&	 5,088,606 	&	 6,017,582\\
0.4 & 39,366 & 42,008&	 12,796 & 15,100&	 214,979 & 405,543&	 3,950,659 	&	 4,856,567\\
0.45 & 36,194 & 38,809&	 10,111 & 12,251&	 190,704 & 381,106&	 3,369,522 	&	 4,261,678\\
0.5 & 34,141 & 36,737&	 8,539 & 10,515&	 182,906 & 373,191&	 3,056,597 	&	 3,938,536\\
0.55 & 29,404 & 31,960&	 6,926 & 8,699&	 110,464 & 299,895&	 1,156,887 	&	 1,918,111\\
0.6 & 23,150 & 25,615&	 5,395 & 7,036&	 84,459 & 272,894&	764,584	&	 1,483,586\\
0.65 & 20,604 & 23,007&	 4,572 & 6,127&	 77,904 & 265,699&	614,743	&	 1,308,939\\
0.7 & 16,391 & 18,707&	 3,929 & 5,386&	 72,651 & 259,877&	509,81	&	 1,182,631\\
0.75 & 15,951 & 18,234&	 3,726 & 5,129&	 71,663 & 258,550&	472,869	&	 1,126,702\\
0.8 & 12,989 & 15,137&	 3,490 & 4,846&	 69,449 & 255,904&	419,533	&	 1,046,786\\
0.85 & 11,533 & 13,530&	 3,313 & 4,568&	 68,555 & 254,703&	391,89	&	 986,811\\
0.9 & 11,053 & 12,976&	 3,214 & 4,437&	 68,186 & 254,138&	377,377	&	 949,637\\
0.95 & 10,875 & 12,756&	 3,105 & 4,290&	 67,871 & 253,623&	369,401	&	 929,765\\
1 & 10,874 & 12,756&	3,079 & 4,250&	 67,798 & 253,390&	367,946	&	 926,380\\
\hline 
\end{tabular} 

\end{table}

\begin{table}[t!]
\caption{Elapsed time for online OA-biclustering}
\centering
		
\begin{tabular}{c|c|c|c|c}

Dataset & $|I|$ & $G|$ & $|M|$ & time, s \\
\hline
co-authoring & 45,904 & 19,885 & 16,400 & 0.13\\

co-occurrence & 183,363 & 13,587 & 9,264 & 0.25\\

actor & 1,470,418 & 127,823 & 383,640 & 3.55\\

p2p & 55,829,392 & 19,86,588 &  5,380,546 & 260.13\\

\end{tabular}
				
\end{table}

An interesting issue has appeared:  At which $\rho_{min}$ the generated biclusters do not cover all formal concepts with non-empty extent and intent? According to our experiments for two-mode (see also Appendix) and one-mode networks, it usually happens around $\rho_{min}=0.5$ or higher (containing intervals marked by two horizontal lines in the tables), so, we may hypothesise that one can normally set minimal density value equal to 0.5.

\subsection{Folksonomies as 3-mode networks}

Folksonomy is a typical example of a three-mode network, where a hyperedge connects a user, a tag, and an attribute. Thus each hyperedge is a set of size three with three vertices of different types; it is convenient to represent edges as tuples $(user, tag, resource)$. Since we experiment with Bibsonomy, a Folksonomy-based resource sharing system  for scientific bibliography, our users are scientists, resources are papers that they bookmarked or even authored; a tag is assigned by a scientist to a particular paper while bookmarking.

\begin{figure}[t]
\begin{center}
	\includegraphics[width=1\textwidth]{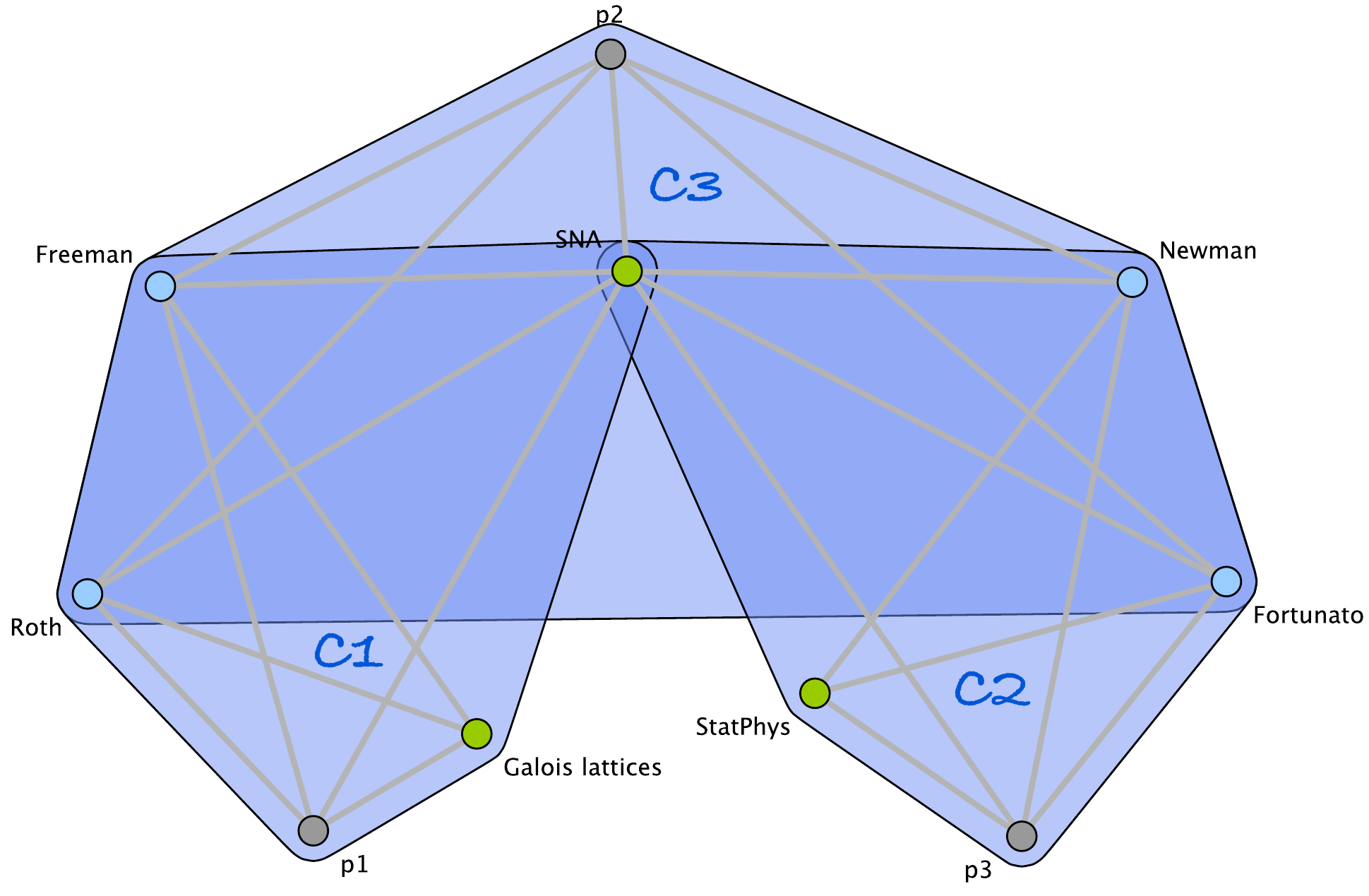}
	\caption{Three triconcepts $C_1$, $C_2$, $C_3$ for the Bibsonomy three-mode network}
	\label{fig:toybib}
\end{center}
\end{figure}

\begin{svgraybox}
Let us consider a toy imaginary example of Bibsonomy data; the input context is shown by three layers in Table~\ref{tbl:toybib}. There are four users ($u_1=Fortunato$, $u_2=Freeman$, $u_3=Newman$, and $u_4=Roth$) and three tags ($t_1=Galois \ Lattices$, $t_2=SNA$, and $t_3=Statistical \ Physics$). Three papers $p_1$, $p_2$, and $p_3$ are marked according to the research interests of those users. Thus Freeman and Roth marked paper 1 by tags ``Galois Lattices'' and ``SNA'', while  Fortunato and Newnam tagged paper 3 by tags `SNA'' and ``Statistical Physics''. All the users assigned tag ``SNA'' to paper 2. Three corresponding communities can be easily captured by formal triconcepts:

$$C_1=(\{u_2, u_4\},\{t_1,t_2\},\{p_1\})$$
$$C_2=(\{u_1, u_3\}, \{t_2, t_3\},\{p_3\})$$
$$C_3=(\{u_1, u_2, u_3, u_4\},\{t_2\},\{p_2\}).$$

Concept $C_3$ is more general than $C_1$ and $C_2$ w.r.t to extent inclusion, and corresponds to SNA-interested users, while $C_1$ corresponds to those, who interested in concept lattices for SNA domain, and $C_2$ unites users interested in SNA by means of methods similar to their prototypes in Statistical Physics. The corresponding hypergraph with these triconcepts is shown in Fig.~\ref{fig:toybib}.

\end{svgraybox}

\begin{table}
\caption{A toy example with Bibsonomy data}\label{tbl:toybib}
\centering 
 \begin{tabular}{ccccc} \begin{tabular}{cccc}

& $t_1$ & $t_2$ &$t_3$\\
\cline{2-4} $u_1$ &\multicolumn{1}{|c|}{} & \multicolumn{1}{|c|}{} &  \multicolumn{1}{|c|}{} \\
\cline{2-4} $u_2$ &  \multicolumn{1}{|c|}{$\times$} &  \multicolumn{1}{|c|}{$\times$} &  \multicolumn{1}{|c|}{} \\ \cline{2-4} $u_3$ &  \multicolumn{1}{|c|}{} &  \multicolumn{1}{|c|}{} &  \multicolumn{1}{|c|}{} \\ \cline{2-4} 
$u_4$ &  \multicolumn{1}{|c|}{$\times$} &  \multicolumn{1}{|c|}{$\times$} &  \multicolumn{1}{|c|}{} \\ \cline{2-4}

& \multicolumn{3}{c}{$p_1$}\\

\end{tabular} & \quad\quad &

\begin{tabular}{cccc}

& $t_1$ & $t_2$ &$t_3$\\
\cline{2-4} $u_1$ &\multicolumn{1}{|c|}{} & \multicolumn{1}{|c|}{$\times$} &  \multicolumn{1}{|c|}{} \\
\cline{2-4} $u_2$ &  \multicolumn{1}{|c|}{} &  \multicolumn{1}{|c|}{$\times$} &  \multicolumn{1}{|c|}{} \\ \cline{2-4} $u_3$ &  \multicolumn{1}{|c|}{} &  \multicolumn{1}{|c|}{$\times$} &  \multicolumn{1}{|c|}{} \\ \cline{2-4} $u_4$ &  \multicolumn{1}{|c|}{} &  \multicolumn{1}{|c|}{$\times$} &  \multicolumn{1}{|c|}{} \\ \cline{2-4}

& \multicolumn{3}{c}{$p_2$}\\

\end{tabular}

& \quad\quad &

\begin{tabular}{cccc}

& $t_1$ & $t_2$ &$t_3$\\
\cline{2-4} $u_1$ &\multicolumn{1}{|c|}{} & \multicolumn{1}{|c|}{$\times$} &  \multicolumn{1}{|c|}{$\times$} \\
\cline{2-4} $u_2$ &  \multicolumn{1}{|c|}{} &  \multicolumn{1}{|c|}{} &  \multicolumn{1}{|c|}{} \\ \cline{2-4} $u_3$ &  \multicolumn{1}{|c|}{} &  \multicolumn{1}{|c|}{$\times$} &  \multicolumn{1}{|c|}{$\times$} \\ \cline{2-4} $u_4$ &  \multicolumn{1}{|c|}{} &  \multicolumn{1}{|c|}{} &  \multicolumn{1}{|c|}{} \\ \cline{2-4}

& \multicolumn{3}{c}{$p_3$}\\

\end{tabular}
\end{tabular}
\end{table}

To build all triconcepts of a certain context we have used a Java implementation of the TRIAS algorithm by R. J{\"a}schke \cite{Jaschke:2006}. The last
two columns in Table \ref{exp1} mean time of execution of TRIAS and OAC-prime algorithms.

\begin{table}[t!]
\caption{Experimental results for $k$ first triples of Bibsonomy data set with $\rho_{min}=0$}
\centering{
\tabcolsep=2pt

\begin{tabular}{ccccccccccc}
\hline
$k$, number of  &  $|U|$ & $|T|$ & $|R|$ & $|\mathfrak{T}|$  &  $|\mathcal{T}_{OAC\prime}|$ &  & TRIAS, s & \multicolumn{2}{c}{OAC-Prime,s }\\
first triples &  &&&&&&& full time & online phase\\
\hline
100    &  1    & 47    & 52    & 57    &  77 &   & 0.2   & 0.02 &  0.003\\
1,000   &  1    & 248   & 482   & 368   & 656  &   & 1     & 0.043     &  0.001 \\
10,000  &  1    & 444   & 5,193  & 733   &  1,461 &   & 2     & 273 & 0.031 \\
100,000 &  59   & 5,823  & 28,920 & 22,804 & 33,172  & & 3,386  & 24,185  & 0.542 \\
200,000 &  340  & 14,982 & 61,568 & -     & 105,571 &  & $>$ 24 h & 25,446 & 1.268 \\
500,000 & 1,191 & 45,232 & 148,695 & - & 316,139 &  & $>$ 24 h  & 29,035 & 3.529\\
816,197 & 2,467 & 69,904 & 268,692 & - & 484,349 &  & $>$ 24 h  & 241,341 & 5.186\\
\hline
\end{tabular}
}
\label{exp1}
\end{table}

Note that here we have reported both the full execution time of OAC-prime algorithm, i.e. tricluster generation with density calculation, and the time of online phase for tricluster generation only. One may note a dramatical drop-off in time efficiency between the last and penultimate lines in Table~\ref{exp1} for the full execution time, while online phase took only about half a second more. The devil is in the hashing datastructures used for duplicate elimination and we believe the timing can be improved, for example by a specially designed Bloom filter.   
Note that a more general and efficient algorithm Data-Peeler \cite{Cerf:2013} could be used suitable for mining $n$-concepts. 

Distribution of density of triclusters for all the triples of Bibsonomy dataset is given in Table \ref{exp2}.

\begin{table}[t!]
\caption{Density distribution of OAC-prime triclusters for 816,197 triples of Bibsonomy data set with $\rho_{min}=0$}
{
\begin{center}
\begin{tabular}{ccc}
\hline
lower bound of $\rho$ & upper bound of $\rho$ & number of triclusters\\
\hline

%0 &	0,05 &	18617\\
%0,05 &	0,1 &	195\\
%0,1	& 0,2 &	112\\
%0,2 & 0,3 &	40\\
%0,3	& 0,4 &	20\\
%0,4	& 0,5 &	10\\
%0,5	& 0,6 &	8\\
%0,6	& 0,7 &	1\\
%0,7	& 0,8 &	1\\
%0,8	& 0,9 &	0\\
%0,9	& 1 & 49\\

0 &	0,05 &	172\\

0,05 &	0,1 &	3,070\\

0,1	& 0,2 &	36,878\\

0,2 & 0,3 &	77,170\\

0,3	& 0,4 &	90,005\\

0,4	& 0,5 &	67,659\\

0,5	& 0,6 &	66,711\\

0,6	& 0,7 &	41,507\\

0,7	& 0,8 &	22,225\\

0,8	& 0,9 &	11,662\\

0,9	& 1 &  67,290\\

\hline
\end{tabular}
\end{center}
}\label{exp2}
\end{table}

\subsection{MovieLens data as 4-mode network}

We summarise the results of prime-based tetraclustering execution on Movielens data below:

%\begin{table}[t!]
%\caption{Statistics for the tetraclusters' collection from Movielens data	}\label{tbl:4clust}
%\centering
\begin{center}
\begin{tabular}{lc}

Time: & 13,252 ms\\

Number of $n$-clusters: & 89,931\\

Average volume, $\overline{Vol}$: & 455.4\\

Average density, $\overline{\rho}$: & 0.35\\

Average coverage: & 0.1\%\\

Average mass, $\overline{mass}$: & 103.7\\

Average $\rho\cdot mass$: & 28.1.\\

\end{tabular}
\end{center}
%\end{table}

In addition to average density we report average volume, average coverage (the number of covered original tuples by each tetracluster on average), average mass (the number of tuples inside each tetraclusters on average), and quite an interesting statistic, average $\rho\cdot mass$. If we maximise the latter criterion, then we require for our tetraclusters to be dense and large at the same time while criterion $\rho\cdot Vol$ could result in sparse patterns.

To provide concrete examples of tetra-clusters, we have selected rather small-sized dense communities in Table~\ref{tbl:4clust}.

\begin{table}[t!]
\caption{Tetraclustres for Movielens data	}\label{tbl:4clust}
\centering
\begin{tabular}{c|c|c|c|c|c|c}
\hline 
no. & Generating tuple & volume &	$\rho$	& coverage	& $mass$	 & $\rho\cdot mass$ \\
\hline 
1 &  (483, Star Trek IV, 5, 1997/11) & 27 & 0.93 & 0.03 \% & 25 & 23.1  \\
2 &  (384, Evita, 5, 1998/03) & 15 & 0.87 & 0.01 \% & 13 & 11.3 \\
3 &  (872, Scream 2, 5, 1998/02) & 15 & 0.87 & 0.01 \% & 13 & 11.3 \\
4 &  (102, Face/Off, 3, 1997/10) & 12 & 0.92 & 0.01 \% & 11 & 10.1 \\
5 &  (750, Gang Related, 1, 1997/11) & 9 & 1.00 & 0.01 \% & 9 & 9.0 \\
\hline 
\end{tabular}

\vspace{0.5cm}

\begin{tabular}{c|c|c|c|c}
\hline 
no. & 	users	& movies &	rating	& time\\
\hline 

1 &   \{109,307,374,483,87,& \{Star Trek: The Wrath of Khan (82), Star Trek IV:& \{5\} & \{97/11\}\\

&   545,815,882,927\} &  The Voyage Home (86), Star Wars (77) \}  &  & \\
\hline 
2 &  \{378,384,392\} & \{Good Will Hunting (97), Evita (96), Titanic (97), & \{5\} & \{98/03\}\\
&  &  L.A. Confidential (97),  As Good As It Gets (97)\} &  & \\
\hline 
3 &   \{206,332,872\} & \{Time to Kill, A (96), Scream (96), Scream 2 (97), & \{5\} & \{98/02\}\\
&  & Air Force One (97), Titanic (97)\} &  & \\
\hline 
4 &   \{102,116,268,430\} & \{Grosse Pointe Blank (1997), Face/Off (1997) \} & \{3\} & \{97/10\}\\
 &  &  Air Force One (1997)\} &  & \\
\hline 
5 &  \{181,451,750\} & \{Gang Related (1997), Rocket Man (1997) & \{1\} & \{97/11\}\\
&  & Leave It to Beaver (1997)\} & & \\
\hline 
\end{tabular}

\end{table}

For example, one can easily identify the community of modern space opera lovers in 4-cluster no.~1. Note that their third and fourth components are always sets containing a single element due to the chosen mode nature: the same people cannot rate the same movies by different marks simultaneously or within a different month.

\subsection{1-mode networks as two-mode ones}

There are different techniques called projections to transform two-mode graphs to their one-mode versions \cite{Latapy:2008,Newman:2001}. Sometimes, researchers even do transformations in backward direction to consider interactions between different subgroups of actors as they were from different modes of the corresponding two-mode network \cite{Doreian:2004,Zakhlebin:2015}.

A undirected one-mode network in the form $\Gamma=(G,E\subseteq G \times G)$ can be considered as the two-mode network by composing a context $\K=(G,G,I)$ where $gEh \iff gIh$ for any $g,h \in G$, with two options for $I$ being a symmetric relation: a) reflexive and b) irreflexive. 

In reflexive case, each concept $(A,B)$ of such context $\K$ that fulfils $A=B$ corresponds to the maximal clique $A$ in the original one-mode network. 

We provide the reader with the results of OA-biclustering for one-mode networks in Tables~\ref{karate}, \ref{florent1}, \ref{florent2}, \ref{hitech}, and \ref{mexican}.

In addition to the fraction of covered concepts by component-wise set inclusion we have reported intervals $[\rho_{\alpha},\rho_{\beta}]$, where the fraction of covered concepts decreases below 1 first time for each dataset (see two vertical lines in the tables).
		
%\begin{table}[t!]
%\caption{Karate club	34x34, 78 edges	}\label{karate}
%\centering
%\begin{tabular}{c|c|c|c|c}
%$\rho$	& concept  &	Unique  & Biclusters	& Fraction of \\
%	&  coverage &	biclusters & & covered concepts\\
%\hline 
%0	 & 24	 & 43	 & 78	 & 1.00\\
%0.05	 & 24	 & 43	 & 78	 & 1.00\\
%0.1	 & 24	 & 43	 & 78	 & 1.00\\
%0.15	 & 24	 & 43	 & 78	 & 1.00\\
%0.2	 & 24	 & 43	 & 78	 & 1.00\\
%0.25	 & 24	 & 42	 & 77	 & 1.00\\
%0.3	 & 24	 & 41	 & 76	 & 1.00\\
%0.35	 & 24	 & 41	 & 76	 & 1.00\\
%0.4	 & 24	 & 40	 & 75	 & 1.00\\
%0.45	 & 24	 & 39	 & 73	 & 1.00\\
%0.5	 & 24	 & 36	 & 69	 & 1.00\\
%\hline
%0.55	 & 24	 & 32	 & 64	 & 1.00\\
%0.6	 & 23	 & 27	 & 59	 & 0.96\\
%\hline
%0.65	 & 22	 & 23	 & 54	 & 0.92\\
%0.7	 & 18	 & 17	 & 48	 & 0.75\\
%0.75	 & 18	 & 16	 & 44	 & 0.75\\
%0.8	 & 15	 & 12	 & 32	 & 0.63\\
%0.85	 & 12	 & 10	 & 29	 & 0.50\\
%0.9	 & 12	 & 10	 & 29	 & 0.50\\
%0.95	 & 10	 & 9	 & 25	 & 0.42\\
%1	 & 8	 & 8	 & 17	 & 0.33\\
%
%\end{tabular}
%				
%\end{table}

\begin{table}[t!]
\caption{Karate club: 34x34, 190 edges}\label{karate}
\centering
\begin{tabular}{c|c|c|c|c}
$\rho$	& Covered  &	Unique  & Biclusters	& Fraction of \\
	&  concepts &	biclusters & 	& covered concepts\\
\hline 
0 & 134 & 190 & 190 & 1,00\\
0,05 & 134 & 190 & 190 & 1,00\\
0,1 & 134 & 190 & 190 & 1,00\\
0,15 & 134 & 190 & 190 & 1,00\\
0,2 & 134 & 190 & 190 & 1,00\\
0,25 & 134 & 190 & 190 & 1,00\\
0,3 & 134 & 184 & 184 & 1,00\\
0,35 & 134 & 178 & 178 & 1,00\\
0,4 & 134 & 163 & 163 & 1,00\\
\hline
0,45 & 134 & 142 & 142 & 1,00\\
0,5 & 132 & 128 & 128 & 0,99\\
\hline
0,55 & 126 & 108 & 108 & 0,94\\
0,6 & 115 & 91 & 91 & 0,86\\
0,65 & 97 & 71 & 71 & 0,72\\
0,7 & 90 & 67 & 67 & 0,67\\
0,75 & 68 & 47 & 47 & 0,51\\
0,8 & 31 & 25 & 25 & 0,23\\
0,85 & 27 & 20 & 20 & 0,20\\
0,9 & 12 & 12 & 12 & 0,09\\
0,95 & 12 & 12 & 12 & 0,09\\
1 & 12 & 12 & 12 & 0,09\\

\end{tabular}
				
\end{table}

In addtion to the reported statistics, let us demonstrate found biclusters and concepts for Zachary's karate club dataset. Originally, the author of \cite{Zachary:1977}, an anthropologist, described social relationships between members of a karate club in the period of 1970--72; the network contains 34 active members of the karate club who interacted outside the club, including 78 pairwise links between them. The club was split into two parts after a conflict between its instructor and president. This dataset is usually used as a benchmark for demonstration and testing of community detection algorithms~\cite{Barabasi:2016}.

\begin{figure}[t]
\begin{center}
	\includegraphics[width=1\textwidth]{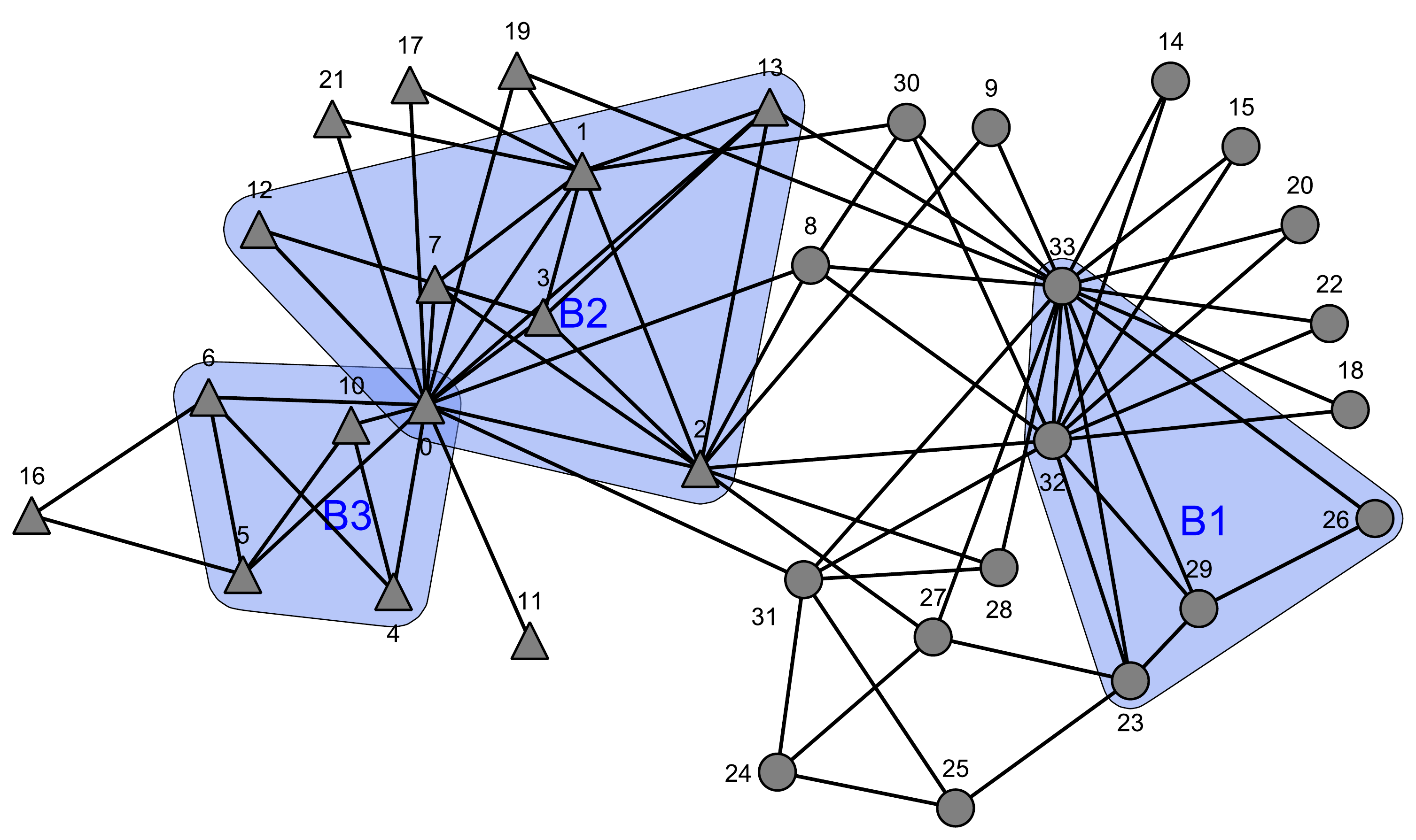}
	\caption{Three dense biclusters $B_1$, $B_2$, $B_2$ found in Karate club network with $\rho_{min}=0.8$}
	\label{karatebicl}
\end{center}
\end{figure}

\begin{svgraybox}
 In Fig.~\ref{karatebicl}, one can see three biclusters ($B_1$, $B_2$, and $B_3$) with density less than 1 but greater 0.8 each.
 Thus none of them is a concept; moreover, union of their intent and extent does not form  a clique of the input one-mode network. 
 
   $$B_1=(29',29')=(\{32,33,26,29,23\}, \{32,33,26,29,23\}) \mbox{ with } \rho=0.84$$
   
   $$B_2=(3',12')=(\{0,1,2,3,7,12,13\}, \{0,3,12\}) \mbox{ with } \rho=0.81$$

   $$B_3=(5',4')=(\{0,10,4,6\}, \{0,10,4,5\}) \mbox{ with } \rho=0.88$$
   
Among all generated concepts, each concept $(X,Y)$ with $X=Y$ results in clique $X$.

Thus concept $(\{0,1,2,3,7\},	\{0,1,2,3,7\})$ forms clique $Q_1=\{0,1,2,3,7\}$, while 
 concepts $(\{0,1,2,3,13\},	\{0,1,2,3,13\})$ and $(\{32,33,29,23\},	\{32,33,29,23\})$ result in $Q_2=\{0,1,2,3,13\}$ and $Q_3=\{32,33,29,23\}$, respectively. Those are cliques of maximal size 5 and 4 from two parts of the karate club after its fission. It is evident that for each of those cliques its set of vertices can be found in some OA-bicluster.
One can check that the set of vertices of $B_1$ contains those of $Q_3$, and vertices of $B_2$ include those of $Q_1$ and $Q_2$. So,  it is possible to conclude that even though the density of a bicluster may be less than 1, they can contain more vertices resulting in larger communities than cliques. 
Note that the club instructor, 0, belongs to extents of $B_2$ and $B_3$ being a ``missing link'' between two corresponding subcommunities, which lack in active interaction otherwise.  

\end{svgraybox}

\begin{figure}[t]
\begin{center}
	\includegraphics[width=1\textwidth]{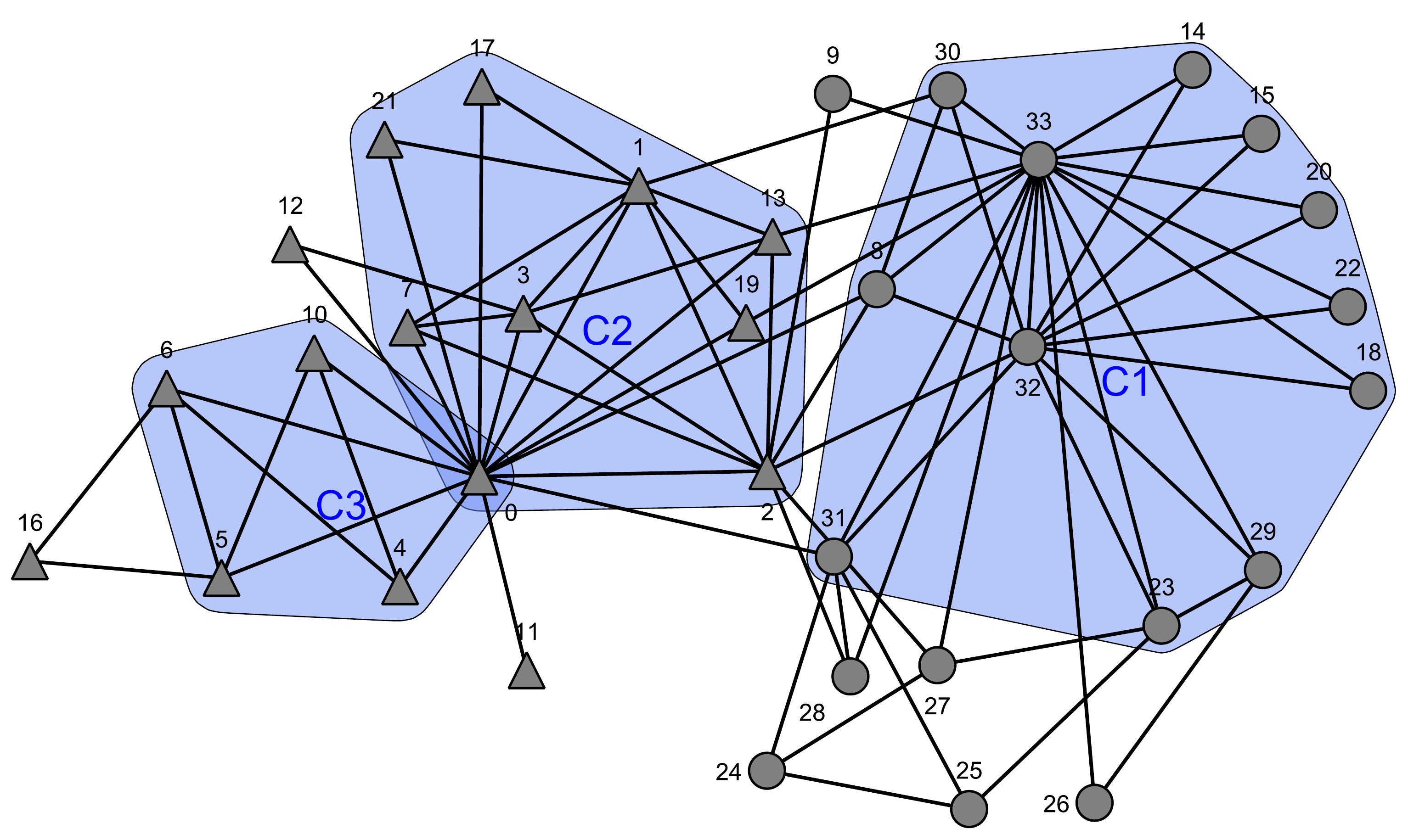}
	\caption{Three formal concepts $C_1$, $C_2$, $C_2$ found in Karate club network}
	\label{karateconc}
\end{center}
\end{figure}

\begin{svgraybox}

 In Fig.~\ref{karateconc}, one can see three found communities that are composed of vertices corresponding to three concepts $C_1$, $C_2$, and $C_3$.

$$C_1=(\{32,33\}, \{32,33,8,14,15,18,20,22,23,29,30,31\})$$

$$C_2=(\{0,1\}, \{0,1,2,3,7,13,17,19,21\})$$

$$C_3=(\{0,10,6\}, \{0,4,5\})$$

In this concrete example, the usage of formal concepts for representing communities seems to be even more beneficial than that of dense OA-biclusters since we have been able to cover almost both parts of the separated karate club by three concepts without sharing members between the counterparts; concepts $C_1$ and $C_2$ contain more vertices than biclusters $B_1$ and $B_2$ shown in Fig~\ref{karatebicl}. Note that the semantic of $C_1$ lies in the interpretation of its intent as common contacts of 32 and 33, an active club member who is loyal to the club's president and the president, respectively. Intent of $C_2$ contains members mutually connected with the club instructor, 0, and member 1.

\end{svgraybox}

\section{Related work}\label{sec:rel}

There is a so-called subspace clustering \cite{Agrawal:2005} closely related to biclustering, where objects are considered as points in high dimensional space and clustered within multidimensional grid of a certain granularity. However, these methods cannot be directly applied to multidimensional relational data, i.e. multi-mode networks, since entities from different modes are often numbered arbitrarily and do not follow a pre-specified order like values along numerical axes. 
However, biclustering of numerical data, which may describe two-mode weighted networks, can be realised with Triadic Concept Analysis in case we consider attribute values as a mode of conditions under which an object has an attribute \cite{Kaytoue:2013}. These results are also applicable to $n$-dimensional numerical datasets. Two other ways to deal with numeric data are to apply the so called scaling, e.g., by a binary threshold, or Pattern Structures defined on vectors of numeric intervals \cite{Ganter:2001,Kaytoue:2011,Codocedo:2014}. Pattern Structures were also used to rethink collaborative filtering and find relevant taste communities for a particular user in terms of vectors of desirable rating intervals for good movies \cite{Ignatov:2015r}.

As for OA-biclustering, it has been used in several applications; for example, OA-biclustering has been applied for finding market segments in two-mode data on Internet advertising to recommend advertising terms to companies playing on these segments \cite{Ignatov:2008,Ignatov:2012a}. In crowdsourcing platforms, OA-biclustering helps to find similar ideas (proposals) to discuss  or potential collaborators\cite{Ignatov:2014b,Ignatov:2014c} as well as answer questions \cite{Chatterjee:2017}; in case we consider opinions of users over a set of different ideas (proposals), it is possible to find antagonists, which may be prospective opponents in crowdsourcing teams \cite{Ignatov:2014a}.   

In fact, biclustering is a well-established tool in Bioinformatics, especially for Gene Expression Analysis in genes-samples networks \cite{Kaytoue:2011,Padilha:2017}. A non-exhaustive concept lattice based taxonomy of biclustering techniques can be found in \cite{Ignatov:2015b}. Methods for three-mode networks are applicable in this domain when in addition to genes and samples time mode comes \cite{Zhao:2005}. 

Going back to networks, several researchers define other kind of networks where the role of dimensions is played by different types of labels of multi-edges between actors \cite{Berlingerio:2013,Berlingerio:2013d}; they call such networks multidimensional while others use the term multi-relational networks \cite{Wu:2015}.

One more variation of networks is realised by $n$-partite networks  where connection are edges between vertices of allowed types \cite{Spyropoulou:2014}. It is possible to mine maximal closed and connected subgraphs in them and interpret them as communities \cite{Lijffijt:2016}; these patterns coincide with bicliques and formal concepts in two-mode case. However, for higher dimensions such $n$-partite graphs are not equivalent to $n$-adic contexts and may result in information loss or phantom hyperedges if we reduce the latter to the former or vice versa \cite{Ignatov:2016}. In \cite{Gnatyshak:2012}, for analysing such tripartite network composed by two two-mode networks with one shared part, biclusters from these two networks have been used. Namely, those biclusters that are similar with respect to their extents are merged by taking the intersection of their extents. The intent of the first bicluster and the intent of the second bicluster become the intent and modus respectively of the resulting tricluster. In FCA domain, analysis of $n$-partite and multi-relational networks can be unified withtin Relational Concept Anlaysis where objects can be invloved in different types of relations with attributes and each other \cite{Hacene:2013}.

Another related subject is tensor factorisation, which is of high importance in Data Mining \cite{Papalexakis:2016} and Machine Learning \cite{Cichocki:2016} due to its ability to reduce data dimensionality, find the so-called hidden factors, and even perform information fusion. The closest approaches to ones in the presented study can be found in works on Boolean matrix \cite{Belohlavek:2010,Belohlavek:2015} and tensor factorisation \cite{Miettinen:2011,Belohlavek:2013}. Thus in \cite{Belohlavek:2010} it was shown that formal concepts  may result in optimal factors in Boolean matrix decomposition; in \cite{Ignatov:2014,Akhmatnurov:2015} these decompositions showed their competitive applicability to collaborative filtering by finding communities of similar tastes. Tensor clustering is another way to find dense patterns; this approach is very similar to multimodal clustering in $n$-ary relations, especially in case of Boolean tensors, which normally represent $n$-ary relations between entities \cite{Ignatov:2011,Mirkin:2011,Metzler:2015,Shin:2016}. An interesting issue here, whether it is possible to obtain improvements in classification accuracy for tensors with labeled objects from one of their dimensions over conventional object-attribute representations \cite{Zhuk:2014}.

Since the proposed multimodal clustering is an approach to find approximate patterns, not absolutely dense as closed $n$-sets or $n$-adic concepts, various similar ideas can be proposed. Thus, in \cite{Cerf:2013} another type of fault-tolerant patterns was proposed, which is guided by the number of allowed non-missing tuples inside an $n$-cluster rather than by  maximising their relative number.  It seems that techniques  searching for relaxed $n$-cliques maximal according a density-like criteria can be proposed for multimode networks as well \cite{Veremyev:2016}. The classic definiton of \emph{biplex} can be compared with the one of OA-bicluster as many more similar relaxations for cliques and their possible $n$-adic generalisations \cite{Borgatti:1997}. 

Comparison of several existing triclustering techniques based on spectral clustering (\textsc{SpecTric}), least squares approximation (\textsc{TriBox}), OAC-prime and OAC-box operators, and formal triconcepts (\textsc{TRIAS}) can be found in \cite{Ignatov:2013,Ignatov:2015}. In \cite{Ignatov:2015}, the complexity of the problem of optimal triclustering cover with respect to several quality criteria is discussed; it is shown that the problem belongs to NP-complete complexity class whereas the problem of the number of such covers belongs to \#P.

Formal concepts and their lattices have been used in criminal studies to find communities of criminals operating together \cite{Poelmans:2012p}. Many more successful applications based on FCA are known as well as related models and techniques \cite{Poelmans:2013a,Poelmans:2013b}. A comprehensive inroduction to FCA can be found in the recent book  \cite{Ganter:2016} and applicaton-oriented tutorial \cite{Ignatov:2014t}.

\section{Conclusions}\label{con}

In fact, we have proposed a scalable technique for community detection in $n$-mode networks (where nodes are normally connected by hyperedges in case of $n>2$). The approach welcomes improvements and may benefit from fine tuning and efficient filtering criteria in order to increase the scalability at the stage of density calculation and guarantee high-quality of the found communities. We consider several directions for such improvements: efficient hashing for elimination of duplicate patterns, strategies for approximate density calculation and selection of meaningful $n$-clusters as well as theoretical justification of choosing good thresholds for minimal density of $n$-clusters.

The proposed technique also can be compared with other exisiting approaches like fault-tolerant $n$-concepts (\cite{Cerf:2013}) and with possible multimodal extensions of the existing ones like different techniques for relaxed cliques  \cite{Veremyev:2016}, variations of bicliques \cite{Nussbaum:2012} or higher-order exentions of modularity-based criteria (\cite{Murata:2010}).
 
Since we have only showcased several relevant examples to community detection in multi-mode networks, validation of the method for analysing similar cases requires domain expert feedback, for example, by a sociologist-practitioner.

\subsubsection*{Acknowledgements.}
We would like to thank our colleagues Rakesh Agrawal, Lo\"{i}c Cerf, Vincent Duquenne, Santo Fortunato, Bernhard Ganter, Jean-Fran\c{c}ois Boulicaut, Mehdi Kaytoue, Boris Mirkin, Amedeo Napoli, Lhouri Nourine, Engelbert Mephu-Nguifo, Sergei Kuznetsov, Rokia Missaoui, Sergei Obiedkov, Camille Roth, Takeaki Uno, Stanley Wasserman, and Leonid Zhukov for their inspirational discussions or a piece of advice, which directly or implicitly influenced this study. We are grateful to our colleagues from the Laboratory for Internet Studies for their piece of advice as well. The study was implemented in the framework of the  Basic Research Program at the National Research University Higher School of Economics in 2016 and 2017 and in the Laboratory of Intelligent Systems and Structural Analysis. The first author has also been supported by Russian Foundation for Basic Research.

\bibliographystyle{spmpsci}

\bibliography{bib}

\clearpage

\section*{Appendix. Experiments with one-mode networks}

%\subsection*{Appendix 1. Experiments with one-mode networks}

\begin{table}[ht!]
\caption{Florent family 1: 16x16, 58 edges}\label{florent1}
\centering
\begin{tabular}{c|c|c|c|c}
$\rho$	& Covered  &	Unique  & Biclusters	& Fraction of \\
	&  concepts &	biclusters & 	& covered concepts\\
\hline 	
0& 43& 58& 58& 1,00\\
0,05& 43& 58& 58& 1,00\\
0,1& 43& 58& 58& 1,00\\
0,15& 43& 58& 58& 1,00\\
0,2& 43& 58& 58& 1,00\\
0,25& 43& 58& 58& 1,00\\
0,3& 43& 58& 58& 1,00\\
0,35& 43& 58& 58& 1,00\\
0,4& 43& 57& 57& 1,00\\
0,45& 43& 53& 53& 1,00\\
0,5& 43& 47& 47& 1,00\\
\hline
0,55& 43& 40& 40& 1,00\\
0,6& 37& 31& 31& 0,86\\
\hline
0,65& 33& 28& 28& 0,77\\
0,7& 29& 19& 19& 0,67\\
0,75& 29& 19& 19& 0,67\\
0,8& 11& 8& 8& 0,26\\
0,85& 9& 6& 6& 0,21\\
0,9& 5& 5& 5& 0,12\\
0,95& 5& 5& 5& 0,12\\
1& 5& 5& 5& 0,12\\	
\end{tabular}
\end{table}
	
%\begin{table}[t!]
%\caption{Florent family 1		16x16, 40 edges	}\label{florent1}
%\centering
%\begin{tabular}{c|c|c|c|c}
%$\rho$	& concept  &	Unique  & Biclusters	& Fraction of \\
%	&  coverage &	biclusters & 	& covered concepts\\
%\hline 
%0	 & 27	 & 40	 & 40	 & 1.00\\
%0.05	 & 27	 & 40	 & 40	 & 1.00\\
%0.1	 & 27	 & 40	 & 40	 & 1.00\\
%0.15	 & 27	 & 40	 & 40	 & 1.00\\
%0.2	 & 27	 & 40	 & 40	 & 1.00\\
%0.25	 & 27	 & 40	 & 40	 & 1.00\\
%0.3	 & 27	 & 40	 & 40	 & 1.00\\
%0.35	 & 27	 & 40	 & 40	 & 1.00\\
%0.4	 & 27	 & 40	 & 40	 & 1.00\\
%0.45	 & 27	 & 39	 & 39	 & 1.00\\
%0.5	 & 27	 & 39	 & 39	 & 1.00\\
%\hline
%0.55	 & 27	 & 31	 & 31	 & 1.00\\
%0.6	 & 19	 & 15	 & 15	 & 0.70\\
%\hline
%0.65	 & 19	 & 15	 & 15	 & 0.70\\
%0.7	 & 8	 & 8	 & 8	 & 0.30\\
%0.75	 & 8	 & 8	 & 8	 & 0.30\\
%0.8	 & 8	 & 8	 & 8	 & 0.30\\
%0.85	 & 8	 & 8	 & 8	 & 0.30\\
%0.9	 & 8	 & 8	 & 8	 & 0.30\\
%0.95	 & 8	 & 8	 & 8	 & 0.30\\
%1	 & 8	 & 8	 & 8	 & 0.30\\
%
%\end{tabular}
%				
%\end{table}

\begin{table}[t!]
\caption{Florent family 2: 16x16, 46 edges	}\label{florent2}
\centering
\begin{tabular}{c|c|c|c|c}
$\rho$	& Covered  &	Unique  & Biclusters	& Fraction of \\
	&  concepts &	biclusters & 	& covered concepts\\
\hline	
	0 & 27 & 46 & 46 & 1,00\\
0,05 & 27 & 46 & 46 & 1,00\\
0,1 & 27 & 46 & 46 & 1,00\\
0,15 & 27 & 46 & 46 & 1,00\\
0,2 & 27 & 46 & 46 & 1,00\\
0,25 & 27 & 46 & 46 & 1,00\\
0,3 & 27 & 46 & 46 & 1,00\\
0,35 & 27 & 46 & 46 & 1,00\\
0,4 & 27 & 46 & 46 & 1,00\\
0,45 & 27 & 46 & 46 & 1,00\\
0,5 & 27 & 44 & 44 & 1,00\\
0,55 & 27 & 43 & 43 & 1,00\\
0,6 & 27 & 41 & 41 & 1,00\\
\hline
0,65 & 27 & 41 & 41 & 1,00\\
0,7 & 25 & 26 & 26 & 0,93\\
\hline
0,75 & 23 & 22 & 22 & 0,85\\
0,8 & 23 & 19 & 19 & 0,85\\
0,85 & 17 & 14 & 14 & 0,63\\
0,9 & 12 & 12 & 12 & 0,44\\
0,95 & 10 & 10 & 10 & 0,37\\
1 & 10 & 10 & 10 & 0,37\\
	
	\end{tabular}
				
\end{table}
	
%\begin{table}[t!]
%\caption{Florent family 2		16x16, 30 edges	}\label{florent2}
%\centering
%\begin{tabular}{c|c|c|c|c}
%$\rho$	& concept  &	Unique  & Biclusters	& Fraction of \\
%	&  coverage &	biclusters & & covered concepts\\
%\hline 
%0	 & 20	 & 26	 & 30	 & 1.00\\
%0.05	 & 20	 & 26	 & 30	 & 1.00\\
%0.1	 & 20	 & 26	 & 30	 & 1.00\\
%0.15	 & 20	 & 26	 & 30	 & 1.00\\
%0.2	 & 20	 & 26	 & 30	 & 1.00\\
%0.25	 & 20	 & 26	 & 30	 & 1.00\\
%0.3	 & 20	 & 26	 & 30	 & 1.00\\
%0.35	 & 20	 & 26	 & 30	 & 1.00\\
%0.4	 & 20	 & 26	 & 30	 & 1.00\\
%0.45	 & 20	 & 24	 & 28	 & 1.00\\
%0.5	 & 20	 & 24	 & 28	 & 1.00\\
%0.55	 & 20	 & 22	 & 26	 & 1.00\\
%\hline
%0.6	 & 20	 & 12	 & 16	 & 1.00\\
%0.65	 & 16	 & 10	 & 14	 & 0.80\\
%\hline
%0.7	 & 8	 & 6	 & 10	 & 0.40\\
%0.75	 & 8	 & 6	 & 10	 & 0.40\\
%0.8	 & 6	 & 4	 & 8	 & 0.30\\
%0.85	 & 2	 & 2	 & 6	 & 0.10\\plus
%0.9	 & 2	 & 2	 & 6	 & 0.10\\
%0.95	 & 2	 & 2	 & 6	 & 0.10\\
%1	 & 2	 & 2	 & 6	 & 0.10\\
%
%\end{tabular}
%				
%\end{table}

\begin{table}[t!]
\caption{Hi-tech: 36x36, 218 edges}\label{hitech}
\centering
\begin{tabular}{c|c|c|c|c}
$\rho$	& Covered  &	Unique  & Biclusters	& Fraction of \\
	&  concepts &	biclusters & 	& covered concepts\\
\hline 		
0 & 191 & 218 & 218 & 1,00\\
0,05 & 191 & 218 & 218 & 1,00\\
0,1 & 191 & 218 & 218 & 1,00\\
0,15 & 191 & 218 & 218 & 1,00\\
0,2 & 191 & 218 & 218 & 1,00\\
0,25 & 191 & 218 & 218 & 1,00\\
0,3 & 191 & 218 & 218 & 1,00\\
0,35 & 191 & 213 & 213 & 1,00\\
0,4 & 191 & 198 & 198 & 1,00\\
\hline
0,45 & 191 & 174 & 174 & 1,00\\
0,5 & 189 & 134 & 134 & 0,99\\
\hline
0,55 & 163 & 99 & 99 & 0,85\\
0,6 & 126 & 78 & 78 & 0,66\\
0,65 & 86 & 49 & 49 & 0,45\\
0,7 & 65 & 31 & 31 & 0,34\\
0,75 & 47 & 22 & 22 & 0,25\\
0,8 & 28 & 16 & 16 & 0,15\\
0,85 & 16 & 13 & 13 & 0,08\\
0,9 & 16 & 13 & 13 & 0,08\\
0,95 & 12 & 12 & 12 & 0,06\\
1 & 12 & 12 & 12 & 0,06\\
\end{tabular}
				
\end{table}		
		
%\begin{table}[t!]
%\caption{Hi-tech, 36x36, 147 edges		}\label{hitech}
%\centering
%\begin{tabular}{c|c|c|c|c}
%$\rho$	& concept  &	Unique  & Biclusters	& Fraction of \\
%	&  coverage &	biclusters & 	& covered concepts\\
%\hline 
%0	 & 132	 & 146	 & 147	 & 1.00\\
%0.05	 & 132	 & 146	 & 147	 & 1.00\\
%0.1	 & 132	 & 146	 & 147	 & 1.00\\
%0.15	 & 132	 & 146	 & 147	 & 1.00\\
%0.2	 & 132	 & 146	 & 147	 & 1.00\\
%0.25	 & 132	 & 146	 & 147	 & 1.00\\
%0.3	 & 132	 & 145	 & 146	 & 1.00\\
%\hline
%0.35	 & 132	 & 136	 & 137	 & 1.00\\
%0.4	 & 131	 & 117	 & 118	 & 0.99\\
%\hline
%0.45	 & 128	 & 93	 & 94	 & 0.97\\
%0.5	 & 119	 & 72	 & 73	 & 0.90\\
%0.55	 & 102	 & 58	 & 59	 & 0.77\\
%0.6	 & 85	 & 48	 & 49	 & 0.64\\
%0.65	 & 70	 & 35	 & 36	 & 0.53\\
%0.7	 & 41	 & 22	 & 23	 & 0.31\\
%0.75	 & 36	 & 18	 & 19	 & 0.27\\
%0.8	 & 19	 & 12	 & 13	 & 0.14\\plus
%0.85	 & 9	 & 9	 & 10	 & 0.07\\
%0.9	 & 9	 & 9	 & 10	 & 0.07\\
%0.95	 & 9	 & 9	 & 10	 & 0.07\\
%1	 & 9	 & 9	 & 10	 & 0.07\\
%\end{tabular}
%				
%\end{table}

\begin{table}[t!]
\caption{Mexican people: 35x35, 268 edges	}\label{mexican}
\centering
\begin{tabular}{c|c|c|c|c}
$\rho$	& Covered  &	Unique  & Biclusters	& Fraction of \\
	&  concepts &	biclusters & 	& covered concepts\\
\hline 		
0 & 373 & 268 & 268 & 1,00\\
0,05 & 373 & 268 & 268 & 1,00\\
0,1 & 373 & 268 & 268 & 1,00\\
0,15 & 373 & 268 & 268 & 1,00\\
0,2 & 373 & 268 & 268 & 1,00\\
0,25 & 373 & 266 & 266 & 1,00\\
0,3 & 373 & 260 & 260 & 1,00\\
0,35 & 373 & 247 & 247 & 1,00\\
\hline
0,4 & 373 & 225 & 225 & 1,00\\
0,45 & 371 & 189 & 189 & 0,99\\
\hline
0,5 & 360 & 151 & 151 & 0,97\\
0,55 & 348 & 119 & 119 & 0,93\\
0,6 & 298 & 69 & 69 & 0,80\\
0,65 & 211 & 45 & 45 & 0,57\\
0,7 & 141 & 24 & 24 & 0,38\\
0,75 & 86 & 15 & 15 & 0,23\\
0,8 & 17 & 5 & 5 & 0,05\\
0,85 & 13 & 4 & 4 & 0,03\\
0,9 & 1 & 1 & 1 & 0,00\\
0,95 & 1 & 1 & 1 & 0,00\\
1 & 1 & 1 & 1 & 0,00\\	

\end{tabular}
\end{table}

%\begin{table}[t!]
%\caption{Mexican people, 35x35, 117 edges	}\label{mexican}
%\centering
%\begin{tabular}{c|c|c|c|c}
%$\rho$	& concept  &	Unique  & Biclusters	& Fraction of \\
%	&  coverage &	biclusters & 	& covered concepts\\
%\hline 
%0	 & 71	 & 112	 & 117	 & 1.00\\
%0.05	 & 71	 & 112	 & 117	 & 1.00\\
%0.1	 & 71	 & 112	 & 117	 & 1.00\\
%0.15	 & 71	 & 112	 & 117	 & 1.00\\
%0.2	 & 71	 & 112	 & 117	 & 1.00\\
%0.25	 & 71	 & 112	 & 117	 & 1.00\\
%0.3	 & 71	 & 108	 & 113	 & 1.00\\
%0.35	 & 71	 & 104	 & 109	 & 1.00\\
%0.4	 & 71	 & 96	 & 101	 & 1.00\\
%\hline
%0.45	 & 71	 & 89	 & 94	 & 1.00\\
%0.5	 & 70	 & 78	 & 83	 & 0.99\\
%\hline
%0.55	 & 68	 & 63	 & 68	 & 0.96\\
%0.6	 & 62	 & 50	 & 55	 & 0.87\\
%0.65	 & 56	 & 40	 & 45	 & 0.79\\
%0.7	 & 38	 & 28	 & 33	 & 0.54\\
%0.75	 & 34	 & 25	 & 30	 & 0.48\\
%0.8	 & 19	 & 18	 & 23	 & 0.27\\
%0.85	 & 16	 & 15	 & 20	 & 0.23\\
%0.9	 & 13	 & 13	 & 16	 & 0.18\\
%0.95	 & 12	 & 12	 & 13	 & 0.17\\
%1	 & 12	 & 12	 & 13	 & 0.17
%\end{tabular}
%\end{table}

\end{document}